\documentclass[twocolumn,superscriptaddress,amsmath,amssymb,aps,prx]{revtex4-2}
\usepackage{amsmath,amssymb}
\usepackage{graphicx}
\usepackage{dcolumn}
\usepackage{bm}
\usepackage{multirow}
\usepackage{textcomp}
\usepackage{url}
\usepackage{enumitem}
\usepackage{float}
\usepackage[dvipsnames]{xcolor}
\usepackage[colorlinks=true, allcolors=cyan]{hyperref}
\usepackage{orcidlink}

\usepackage{ulem}
\newcommand{\revise}[1]{ {\color{black}#1}}
\begin{document}

\title{Stochastic Similarity Renormalization Group}

\author{R. Z. Hu\,\orcidlink{0009-0002-8797-6622}}
\thanks{These authors contributed equally to this work.}
\affiliation{School of Physics, and State Key Laboratory of Nuclear Physics and Technology, Peking University, Beijing 100871, China}
\author{X. Zhen\,\orcidlink{0009-0000-1806-4123}}
\thanks{These authors contributed equally to this work.}
\affiliation{School of Physics, and State Key Laboratory of Nuclear Physics and Technology, Peking University, Beijing 100871, China}
\author{F. R. Xu\,\orcidlink{0000-0001-6699-0965}}\email[Contact author: ]{frxu@pku.edu.cn}
\affiliation{School of Physics, and State Key Laboratory of Nuclear Physics and Technology, Peking University, Beijing 100871, China}
\affiliation{Southern Center for Nuclear-Science Theory (SCNT), Institute of Modern Physics, Chinese Academy of Sciences, Huizhou 516000, China}
\author{J. C. Pei\,\orcidlink{0000-0002-9286-1304}}\email[Contact author: ]{peij@pku.edu.cn}
\affiliation{School of Physics, and State Key Laboratory of Nuclear Physics and Technology, Peking University, Beijing 100871, China}
\affiliation{Southern Center for Nuclear-Science Theory (SCNT), Institute of Modern Physics, Chinese Academy of Sciences, Huizhou 516000, China}

\date{\today}

\begin{abstract}
By integrating the quantum Monte Carlo technique into the similarity renormalization group (SRG), we have developed a stochastic SRG framework (SRGQMC) capable of both free-space and in-medium many-body evolutions. This approach circumvents the rapid tensor-space explosion of many-body flow equations by mapping continuous unitary transformations onto an ensemble of signed random walkers. We benchmark the SRGQMC against deterministic free-space SRG evolutions of realistic nucleon-nucleon and three-nucleon interactions, as well as against in-medium SRG (IMSRG) many-body calculations with the Richardson pairing model at two- and three-body levels [IMSRG(2)/(3)]. While a deterministic extension to the four-body level [IMSRG(4)] remains unfeasible due to prohibitive computational costs, we have achieved the first IMSRG(4) calculation by using the stochastic technique, demonstrating a substantial improvement toward the full configuration-interaction limit. This stochastic framework provides a practical pathway to higher-order IMSRG calculations.
\end{abstract}

\maketitle

\textit{Introduction.\textemdash}
The renormalization group (RG)~\cite{PhysRevD.48.5863,Wegner2001} provides a powerful reorganizing framework for the descriptions of multi-scale quantum systems, decoupling low-energy physics from high-energy or short-range correlations. The similarity renormalization group (SRG)~\cite{PhysRevC.75.061001,Furnstahl2013} realizes this through a continuous unitary evolution of the Hamiltonian toward a band- or block-diagonal structure while preserving physical observables. The free-space SRG transformation softens nuclear forces through non-perturbative decoupling of high- and low-momentum components, thereby greatly accelerating the convergence of subsequent many-body calculations~\cite{PhysRevLett.107.072501}. The in-medium SRG (IMSRG)~\cite{PhysRevLett.106.222502,HERGERT2016165,ZHEN2025139350} generalizes the evolution strategy of the flow equation, directly evolving the many-body Hamiltonian to decouple the target state of the finite nucleus or infinite nuclear matter.


\revise{The IMSRG has been widely successful in \textit{ab initio} many-body calculations of nuclei, e.g., in the prediction of nuclear driplines~\cite{PhysRevLett.126.022501}, nuclear neutron skins~\cite{Hu2022} and neutrinoless double-beta decays~\cite{PhysRevLett.132.182502,PhysRevLett.124.232501,d2n7-n2c7}. However, the predictive accuracy is restricted by mandatory truncations of induced high-rank terms in Hamiltonian and other operators. The commonly used normal-ordered two-body approximation, IMSRG(2), omits residual three- and higher-body operators. The high-order contributions can be important in descriptions of nuclear observables, such as $2^+$ excitation energies~\cite{PhysRevC.96.014303}, dipole polarizabilities~\cite{PhysRevC.98.014324,PhysRevLett.124.132502} and $\beta$-decay matrix elements~\cite{PhysRevLett.126.182502}. Previous works estimated that IMSRG(2) has an error of $2\%$--$5\%$ in the correlation energy and an error of $1\%$--$1.5\%$ in the charge radius ~\cite{PhysRevC.103.044318,PhysRevC.111.034311}. High-order correlations also play an important role in nuclear collectivity~\cite{PhysRevC.105.L061303,PhysRevX.15.011028,PhysRevLett.111.252501,PhysRevLett.124.042501} and clustering~\cite{PhysRevLett.106.192501,PhysRevLett.134.162503,Shen2023}.}

\revise{Access to higher orders of truncations is also essential for evaluating many-body uncertainties~\cite{PhysRevC.103.044318,PhysRevC.110.044316,PhysRevC.110.044317,niu2026}. Comparison between successive truncation orders can offer a direct assessment of the uncertainty from omitted correlations. Broader efforts on quantifying the uncertainty are going on~\cite{PhysRevLett.132.182502,y8kt-mgf5}. Precision predictions on the neutrinoless double-beta decay are desired~\cite{PhysRevLett.126.182502,PhysRevLett.126.042502}. However, the rapid growth in the computational cost (including storage) of deterministic tensor-contraction calculations severely restricts access to higher orders of truncations in realistic calculations~\cite{PhysRevC.110.044316,PhysRevC.110.044317}. Low-rank approximations based on randomized singular-value decomposition offer a route to reducing the storage and computational demands of three-body interactions~\cite{TICHAI2021136623,PhysRevResearch.6.043331}. Such approach exploits a compact operator representation, whereas in this work we use a quantum Monte Carlo technique to directly sample the commutators which drive the flow.}

In this Letter, we introduce a stochastic similarity renormalization group with a combination of the quantum Monte Carlo (SRGQMC), in which the continuous unitary flow is represented by a dynamic ensemble of random walkers rather than by the deterministic evolution of operator matrices. The walkers sample the commutator structures of the flow equation and faithfully preserve the renormalization trajectory with controllable statistical uncertainties. In this way, SRGQMC preserves the decoupling principle of SRG while offering a scalable Monte Carlo route toward higher-order IMSRG calculations. The resulting energies calculated at the four-body level show a substantial restoration of missing higher-order correlations toward the full configuration-interaction (FCI) limit, demonstrating that walker sampling can overcome the four-body operator barrier that has traditionally restricted deterministic IMSRG calculations.

\begin{figure*}[t]
    \centering
    \includegraphics[width=1\textwidth]{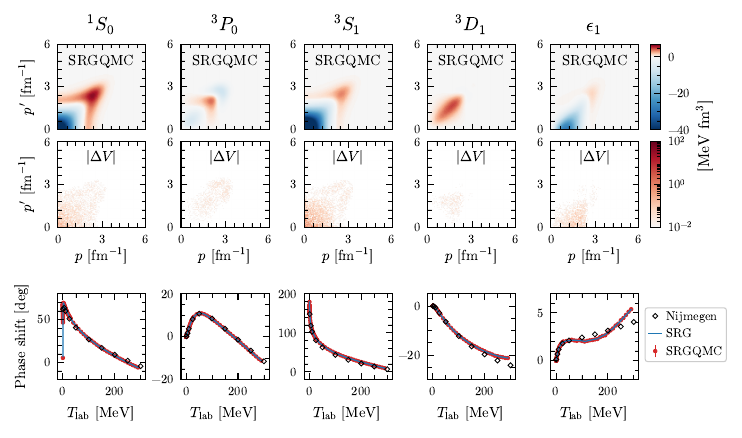}
    \caption{ SRG- and SRGQMC-evolved neutron-proton interactions at $\lambda=2.0\,\mathrm{fm}^{-1}$ for N$^3$LO$_{\rm EMN}(500)$. Columns from left to right denote the $^1S_0$, $^3P_0$, $^3S_1$, $^3D_1$, and $\epsilon_1$ channels. Row 1 presents the loop-averaged SRGQMC matrix elements $V(p',p)$ with $N_\mathrm{w}=10^4$ and $N_{\rm loop}=10$. Row 2 shows their absolute differences from the deterministic SRG results, $|\Delta V|=|V_{\rm SRGQMC}-V_{\rm SRG}|$. Row 3 shows the scattering phase shifts (including the mixing angle $\epsilon_1$) against the laboratory energy $T_{\rm lab}$, derived from SRG (blue solid lines) and SRGQMC (red dots with error bars), compared with the empirical Nijmegen analysis (open diamonds)~\cite{PhysRevC.41.1435} as a reference.}
    \label{fig:nn}
\end{figure*}

\textit{For evolution of NN interaction.\textemdash}
We first formulate the stochastic free-space SRG evolution for a two-nucleon (NN) Hamiltonian. For a given partial wave, the intrinsic Hamiltonian is partitioned as $H(s)=T+V(s)$, where $s$ is the continuous flow parameter, and $T$ is the relative kinetic energy which is $s$-independent. Employing the standard Wegner generator~\cite{Wegner2001}, $\eta(s)=[T,H(s)]$, the unitary evolution is governed by the operator flow equation~\cite{PhysRevC.75.061001}
\begin{equation}
    \frac{{\rm d}H(s)}{{\rm d}s}=[\eta(s),H(s)]
    =[[T,H(s)],H(s)],
    \label{eq:srg_flow}
\end{equation}
which can be written as a set of coupled nonlinear matrix equations on a discrete momentum mesh~\cite{PhysRevC.75.061001},
\begin{equation}
\label{eq:srg_matrix_flow}
    \frac{{\rm d}H_{ij}}{{\rm d}s}
    = \sum_k \left(\eta_{ik}H_{kj}-H_{ik}\eta_{kj}\right),
\end{equation} 
with
\begin{equation}
\eta_{ij}=(\epsilon_i-\epsilon_j)H_{ij},
\end{equation}
where $\epsilon_i$ is the discrete kinetic energy.

In the SRGQMC framework, the deterministic matrix $H_{ij}$ is mapped onto an ensemble of signed discrete walkers residing on all possible matrix sites $(i,j)$. The stochastic estimator of the Hamiltonian is given by $H_{ij}\simeq \mathcal{N} C_{ij}$, where $C_{ij}$ represents the net walker population on site $(i,j)$. To control the simulation size, the scaling factor $\mathcal{N}$ is uniquely defined by the target total walker population $N_\mathrm{w}$ at $s=0$ via
\begin{equation}
    \mathcal{N} = \frac{1}{N_\mathrm{w}} \sum_{ij} |H_{ij}(0)|.
    \label{eq:normalization_factor}
\end{equation}
The walker distribution is initialized in proportion to the unevolved Hamiltonian, $C_{ij}(0)\propto H_{ij}(0)$.


\revise{The stochastic flow is driven by sampling the commutators in Eq.~\eqref{eq:srg_matrix_flow}. During a flow step $\Delta s$, a walker of signed weight $w_{ij}$ at site $(i,j)$ generates contributions through two spawning channels: a left-index update, $(i,j)\rightarrow(k,j)$, and a right-index update, $(i,j)\rightarrow(i,l)$. For the two-body SRG considered here, we sample one target index for each channel uniformly from the allowed indices. For a matrix of dimension $d$, excluding $k=i$ and $l=j$ where the corresponding generator elements vanish gives the proposal probabilities $P_L(k|i)=P_R(l|j)=1/(d-1)$. Once the target indices are selected, the signed contributions added to the target populations are
\begin{align}
    \delta C_{kj} &= \frac{\Delta s}{P_L(k|i)}\,\eta_{ki}w_{ij}, \\
    \delta C_{il} &= -\frac{\Delta s}{P_R(l|j)}\,\eta_{jl}w_{ij}.
    \label{eq:walker_spawn}
\end{align}
For integer walkers, we have $w_{ij}=\pm1$; but more generally, walkers may carry noninteger signed weights with $w_{ij}$ represented as floating-point numbers. The proposal probabilities determine where a contribution is generated, whereas the expressions above determine the magnitude and sign. Since a target is selected with probability $P$ ($P_L$ or $P_R$), its contribution is multiplied by $1/P$ to ensure that the average contribution equals the corresponding term in the flow equation. These contributions are accumulated directly without a Metropolis--Hastings acceptance step. More details can be found in the Supplemental Material~\cite{SM}.}

Each spawning sweep is immediately followed by an annihilation step, where walkers arriving at the same site cancel out via algebraic summation of their signed populations. To guarantee physical consistency, Hermiticity is explicitly enforced at each step by symmetrizing the walker matrix, after which the generator $\eta$ is dynamically updated using the new stochastic Hamiltonian. This sequence delivers a stochastic realization of the continuous SRG flow while bypassing full deterministic matrix-matrix multiplications.

A full stochastic trajectory from $s=0$ to the target resolution scale $\lambda=s^{-1/4}$ defines a single loop. \revise{To reduce statistical uncertainty and estimate sampling errors}, the evolution is repeated across $N_{\rm loop}$ independent loops with distinct random seed series, with statistical uncertainties quoted as standard errors of the mean. Details of uncertainty analysis are given in the Supplemental Material~\cite{SM}.

\revise{SRGQMC shares the stochastic propagation strategy of projector quantum Monte Carlo methods such as Green's function Monte Carlo (GFMC)~\cite{RevModPhys.87.1067,PhysRevLett.113.192501}, while its spawning and annihilation algorithms are similar to those in full configuration-interaction quantum Monte Carlo (FCIQMC)~\cite{Booth2009,q4p2-r1bt,companion_prc}. In GFMC and FCIQMC, walkers sample the many-body wave function, which is propagated in imaginary time toward the ground state of a fixed Hamiltonian. In SRGQMC, walkers instead occupy discrete matrix-element sites, and represent operator coefficients. Their updates sample the commutator structures in the flow equation, and directly decouple the Hamiltonian.}

To illustrate the practical capability of this stochastic framework, we apply SRGQMC to the free-space evolution of the N$^3$LO$_{\rm EMN}(500)$ interaction~\cite{PhysRevC.96.024004} down to a resolution scale of $\lambda=2.0\,\mathrm{fm}^{-1}$. Figure~\ref{fig:nn} presents the evolved interaction matrix elements $V(p',p)$ and the resulting phase shifts, with a walker population of $N_\mathrm{w}=10^4$ and $N_{\rm loop}=10$ independent loops. The first row shows the loop-averaged SRGQMC-evolved matrix elements for the $^1S_0$, $^3P_0$, and coupled $^3S_1$-$^3D_1$ channels, while the second row shows their absolute differences from the deterministic SRG results. The excellent agreement between the interaction matrix elements evolved via SRG and SRGQMC clearly demonstrates that the stochastic dynamics faithfully achieves the objective of the SRG evolution, which is to suppress off-diagonal contributions.

\begin{figure}[t]
\centering
\resizebox{0.45\textwidth}{!}{
\includegraphics{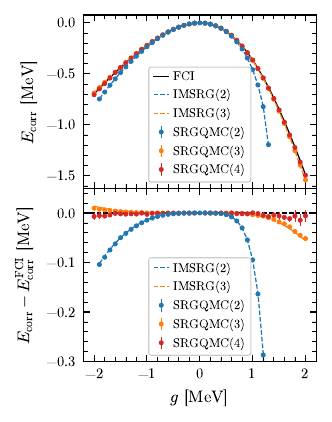}}
\caption{Correlation energy of the $A=4$ Richardson pairing Hamiltonian as a function of the pairing strength $g$. The deterministic IMSRG(2), IMSRG(3) and FCI results are compared with SRGQMC(2), SRGQMC(3) and SRGQMC(4) calculations. Error bars denote the statistical uncertainties from independent stochastic loops. The FCI result is obtained by directly diagonalizing the Hamiltonian in the specified model space.}
\label{fig:pairing}
\end{figure}

A more sensitive validation is provided by the phase shifts shown in the third row of Fig.~\ref{fig:nn}, where our results are also compared with the empirical Nijmegen partial-wave analysis~\cite{PhysRevC.41.1435} as a reference. Since the continuous flow is strictly unitary, these scattering observables serve as a precise diagnosis to verify that the stochastic integration preserves the underlying physics. Quantitatively, the relative root-mean-square (RMS) deviations of the matrix elements between the stochastic and deterministic trajectories remain within $0.5\%$--$5.0\%$ across all the five channels. The minor phase-shift deviations (with the RMS differences below $0.47^\circ$) are also largely encompassed by the quantified statistical error bars. The residual stochastic fluctuations can be systematically suppressed by increasing $N_\mathrm{w}$ or $N_{\rm loop}$.

\revise{It has been well known that the three-nucleon (3N) interaction plays an important role in the descriptions of properties of finite nuclei and infinite nuclear matter. In the present work, we have also validated SRGQMC for the evolution of free-space 3N interactions. The details can be found in the~\hyperref[sec:end_matter]{End Matter}.}

\textit{For many-body calculations.\textemdash}
We have tested many-body calculations using the Richardson pairing model~\cite{RICHARDSON1963277,RevModPhys.76.643,PhysRevLett.134.182502}
\begin{equation}
    \hat{H} = \delta \sum_{p=1}^{p_\mathrm{max}} \sum_{\sigma=\uparrow,\downarrow} (p-1)a_{p\sigma}^\dagger a_{p\sigma} -\frac{g}{2} \sum_{p,q=1}^{p_\mathrm{max}} a_{p\uparrow}^\dagger a_{p\downarrow}^\dagger a_{q\downarrow}a_{q\uparrow},
    \label{eq:pairing_hamiltonian}
\end{equation}
where $\delta$ is the single-particle level spacing (set to $1.0\,\mathrm{MeV}$) and $g$ represents the pairing strength. Although simplified, this model allows us to analyze how the calculation responds to the change in interaction strength. We evaluated a half-filled system with $A = 4$ nucleons distributed across $p_\mathrm{max} = 4$ doubly-degenerate levels, varying $g$ from $-2.0$ to $2.0\,\mathrm{MeV}$. Results are reported in terms of the correlation energy, $E_{\rm corr}$, which is obtained by subtracting the reference state energy from the total energy of the system.

In the IMSRG framework, the many-body Hamiltonian is normal-ordered with respect to a reference state, and truncated to control the operator hierarchy~\cite{HERGERT2016165}
\begin{equation}
\begin{aligned}
    H &= E_0 + \sum_{ij} f_{ij}\{a_i^\dagger a_j\} + \frac{1}{(2!)^2}\sum_{ijkl}\Gamma_{ijkl} \{a_i^\dagger a_j^\dagger a_l a_k\}\\
    &+\frac{1}{(3!)^2}\sum_{ijklmn}W_{ijklmn} \{a_i^\dagger a_j^\dagger a_k^\dagger a_n a_m a_l\}\\
    &+\frac{1}{(4!)^2}\sum_{ijklmnop}X_{ijklmnop} \{a_i^\dagger a_j^\dagger a_k^\dagger a_l^\dagger a_p a_o a_n a_m\}\\
    &+\cdots.
    \label{eq:imsrg2_hamiltonian}
\end{aligned}
\end{equation}
The standard IMSRG(2) truncation restricts the hierarchy to the normal-ordered two-body operator $\Gamma$, while IMSRG(3) and IMSRG(4) further retain the three-body ($W$) and four-body ($X$) tensors, respectively. The flow equation ${\rm d}H/{\rm d}s = [\eta, H]$ then yields a set of coupled equations of tensor contractions for these operators~\cite{SM}.

The core strategy of SRGQMC scales naturally to this many-body tensor space. Instead of tracking full matrices, walker sites are extended to multi-index configurations that identify both the operator rank and tensor indices ($E_0$, $f_{ij}$, $\Gamma_{ijkl}$, $W_{ijklmn}$ or $X_{ijklmnop}$). Rather than calculating deterministic multi-loop sums over intermediate states, each diagrammatic contraction in the flow equation is converted into a corresponding stochastic sampling algorithm. All details of the deterministic flow equations and corresponding sampling algorithms can be found in the Supplemental Material~\cite{SM}. The walker annihilation and loop-averaging algorithms are the same as those in the free-space case.

Figure~\ref{fig:pairing} presents the calculated correlation energies $E_{\rm corr}$ as a function of the coupling strength $g$. SRGQMC($n$) stands for the calculation carried out at the level of $n$-body operators. Using a population of $N_\mathrm{w} = 10^4$ walkers, SRGQMC(2) and SRGQMC(3) calculations reproduce their corresponding deterministic IMSRG results within the quantified statistical uncertainties with relative RMS deviations of $0.35\%$ and $0.51\%$ for the two- and three-body truncations. Figure~\ref{fig:pairing} also illustrates the growing importance of higher-order correlations, omitted in standard IMSRG truncations, within the strongly correlated regime at larger coupling strengths. For $|g|\gtrsim 1.0\,\mathrm{MeV}$, IMSRG(2) rapidly deviates from the exact FCI curve, while the truncation at three-body level provides a remarkable improvement toward the exact solution. The inclusion of four-body operators through SRGQMC(4) (with $N_\mathrm{w} = 10^6$) gives further improvement, well reproducing the FCI result. The comparisons indicate that three- and four-body correlations would play non-negligible roles in the descriptions of strongly correlated systems.

\begin{figure}[t]
\centering
\resizebox{0.43\textwidth}{!}{
\includegraphics{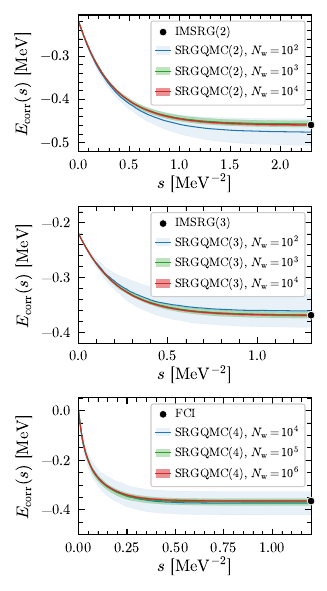}}
\caption{SRGQMC flow of the correlation energy for the $A=4$ Richardson pairing Hamiltonian at $g=1.0\,\mathrm{MeV}$. Shaded bands indicate statistical uncertainties. Deterministic IMSRG(2), IMSRG(3) and FCI results are shown on the right axis by black dots.}
\label{fig:pairing_flow}
\end{figure}

To provide an intuitive and dynamical picture of the stochastic behavior, Fig.~\ref{fig:pairing_flow} details the evolution of the correlation energy as a function of the flow parameter $s$ at $g=1.0\,\mathrm{MeV}$ with different walker populations. We see that the loop-averaged stochastic trajectories at two- and three-body levels smoothly track toward their corresponding exact deterministic calculations as the flow progresses. Deterministic IMSRG(4) has not been available. We benchmark SRGQMC(4) against the exact FCI result which can be obtained for the $A=4$ system by directly diagonalizing the Hamiltonian.
\revise{As expected, increasing the walker population significantly reduces statistical fluctuations. The agreement with deterministic results within the uncertainty bands demonstrates the accuracy attained in these calculations.}

\revise{Due to the nonlinearity of the flow equation, the SRGQMC calculation carries systematic finite-walker bias that is small but cannot be removed by averaging over independent loops. By suggesting a replica method, we have successfully suppressed the bias below the statistical resolution in the present tests. These are detailed in the \hyperref[sec:end_matter]{End Matter} and Supplemental Material~\cite{SM}.}


\revise{SRGQMC offers significant computational advantages over conventional deterministic IMSRG. For example, at a $2\%$ target accuracy of the correlation energy, our test for SRGQMC(3) at $g=0.5\,\mathrm{MeV}$ gives an effective scaling of $M^{3.4(2)}$ (where $M$ is the single-particle basis size), while the deterministic IMSRG(3) has a scaling of $M^9$. A semi-stochastic strategy yields a further speedup by a factor of $1.5$--$2.2$. The more details of the scaling analysis, hybrid approach and memory advantage are discussed in the \hyperref[sec:end_matter]{End Matter} and Supplemental Material~\cite{SM}.}

\textit{Summary and outlook.\textemdash}
In this Letter, we introduce the SRGQMC as a stochastic realization of continuous similarity renormalization group flows, replacing the explicit deterministic evolution of operator matrices with a walker sampling of the same commutator structures. This method preserves the decoupling behavior of SRG while providing fully quantified and controllable statistical uncertainties for the evolved operators and observables. We first validated this idea for the free-space SRG evolution of realistic nucleon-nucleon and three-nucleon interactions, where SRGQMC faithfully reproduces evolved matrix elements. We then benchmarked it against IMSRG(2), IMSRG(3), and full configuration-interaction calculations of the $A=4$ Richardson pairing model, demonstrating that stochastic trajectories exhibit excellent agreement with the deterministic many-body calculations across the entire range of coupling strengths.


The central implication of this work is that SRGQMC provides a highly practical and scalable route to pushing \textit{ab initio} IMSRG calculations toward higher truncation orders.
Accessing higher-order correlations is crucial for high-density nuclear matter~\cite{q4p2-r1bt}, atomic nuclei, and uncertainty quantification of many-body methods~\cite{Booth2009,Booth2012n,PhysRevLett.105.263004,PhysRevLett.134.182502,q3vn-8y8s}.
\revise{While the present work is the first step of SRGQMC many-body calculations with a simplified Hamiltonian, our long-term goal is to apply SRGQMC to higher-order IMSRG calculations using realistic nuclear forces, including stochastic Magnus formulation~\cite{PhysRevC.92.034331,PhysRevC.96.034324}, the  evolution of other observables (including tensor operators), and extension to valence-space IMSRG or multi-reference IMSRG~\cite{PhysRevC.85.061304,PhysRevC.90.041302}. 
}

\textit{Acknowledgments.\textemdash}
This work has been supported by the National Key R\&D Program of China under Grants Nos. 2024YFA1610900 and 2023YFA1606400; the National Natural Science Foundation of China under Grants Nos. 12335007, 12535008 and 12475118; and the High-Performance Computing Platform of Peking University.

\textit{Data availability.\textemdash}
The data that support the findings of this article are openly available~\cite{data}.

\bibliographystyle{modified-apsrev4-2.bst}
\bibliography{reference}

\clearpage

\section*{End Matter}
\label{sec:end_matter}

\begin{figure*}[t]
    \centering
    \includegraphics[width=0.8\textwidth]{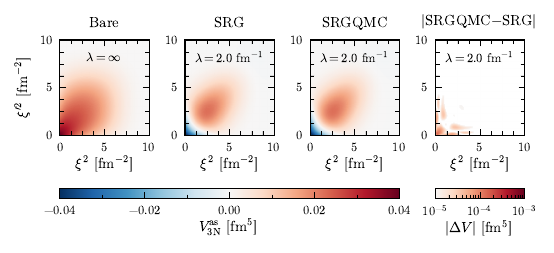}
    \caption{Free-space evolution of the antisymmetrized 3N interaction ($V_{\mathrm{3N}}^{\mathrm{as}}$) for the $\mathcal J^\pi=\tfrac12^+$, $\mathcal T=\tfrac12$ partial wave. Only the $\alpha'=\alpha=0$ channel at $\tan\theta=\tan\theta'=\pi/4$ (see text for definition) is shown. From left to right are the bare interaction, the deterministic SRG at $\lambda=2.0\,\mathrm{fm}^{-1}$, the loop-averaged SRGQMC with $N_\mathrm{w}=10^7$ and $N_{\rm loop}=10$, and the absolute difference $|\Delta V|$ between SRGQMC and SRG. 
    }
    \label{fig:3n}
\end{figure*}

\revise{
\textit{Free-space SRG evolution of three-nucleon interactions.\textemdash}
We extend the free-space SRG calculation to a Hamiltonian containing both NN and three-nucleon (3N) interactions. The input consists of the EMN N$^2$LO(500) NN potential~\cite{PhysRevC.96.024004} and the N$^2$LO 3N force~\cite{HEBELER20211} with $c_D=-1.5$ and $c_E=-0.61$ at the cutoff $\Lambda=500\,\mathrm{MeV}$ using a nonlocal momentum-space regularization.

The SRG flow equation for the 3N interaction is~\cite{PhysRevC.85.021002}
\begin{equation}
\begin{aligned}
    \frac{dV_{123}}{ds}={}&\mathcal{F}_{\mathrm{NN}}
    +[[T_{\mathrm{rel}},V_{\mathrm{NN}}],V_{123}]\\
    &+[[T_{\mathrm{rel}},V_{123}],T_{\mathrm{rel}}+V_{\mathrm{NN}}+V_{123}],
\end{aligned}
\label{eq:free_space_3n_flow}
\end{equation}
where $\mathcal{F}_{\mathrm{NN}}=\sum_{a\ne b}[[T_{\mathrm{rel}},V_a],V_b]$ and $a,b\in\{12,23,31\}$. Here $T_{\mathrm{rel}}$ is the intrinsic three-body kinetic energy and $V_{\mathrm{NN}}=V_{12}+V_{23}+V_{31}$. The source term $\mathcal{F}_{\mathrm{NN}}$ generates an induced three-body interaction even when the initial Hamiltonian contains only NN interactions. Each NN interaction is evolved simultaneously through $dV_a/ds=[[T_a,V_a],T_a+V_a]$, with $T_a$ the relative kinetic energy of pair $a$. In our present implementation, the NN evolution is evaluated deterministically, while the terms involving $V_{123}$ are sampled stochastically.

The calculation uses Jacobi momenta $\mathbf p=(\mathbf k_1-\mathbf k_2)/2$ and $\mathbf q=\tfrac23[\mathbf k_3-(\mathbf k_1+\mathbf k_2)/2]$, and the angular-momentum coupled basis $|pq;[(L_{12}S_{12})J_{12}(l_3\tfrac12)j_3]\mathcal J,(T_{12}\tfrac12)\mathcal T\rangle$. Here $L_{12}$, $S_{12}$, $J_{12}$, and $T_{12}$ describe the pair, while $l_3$ and $j_3$ are the spectator orbital and total angular momenta. The calculation is performed in the $\mathcal J^\pi=\tfrac12^+$, $\mathcal T=\tfrac12$ partial wave with $J_{12}\leq5$. For visualization we define the hypermomentum $\xi$ and hyperangle $\theta$ by
\begin{equation}
    \xi^2=p^2+\frac34q^2,\qquad
    \tan\theta=\frac{p}{\sqrt{3}q/2}.
\end{equation}
Fig.~\ref{fig:3n} displays the bare ($\lambda=\infty$) and evolved ($\lambda=2.0\,\mathrm{fm}^{-1}$) antisymmetrized 3N interaction in the diagonal channel $\alpha'=\alpha=0$, with $(L_{12},S_{12},J_{12},T_{12},l_3,j_3)=(0,0,0,1,0,\tfrac12)$ and $\tan\theta=\tan\theta'=\pi/4$.

The SRGQMC calculation uses $N_\mathrm{w}=10^7$ and $N_{\rm loop}=10$. Fig.~\ref{fig:3n} shows that SRGQMC successfully reproduces the deterministic SRG result, with relative RMS deviation of the matrix elements being 0.14\%. The largest absolute matrix-element deviation is $1.6\times10^{-4}\,\mathrm{fm}^5$. See Supplemental Material~\cite{SM} for $N_\mathrm{w}$- and $N_{\rm loop}$-convergence tests.

\textit{Finite-walker bias and replica correction.\textemdash}
Because the generator is constructed from the stochastically sampled Hamiltonian, the nonlinear SRGQMC flow generally satisfies
$
    \left\langle [\eta(H),H] \right\rangle
    \neq
    [\eta(\langle H\rangle),\langle H\rangle],
$
which produces a systematic finite-walker bias that is small in our present calculations but cannot be removed by averaging over independent loops. We address this issue with the replica method. Two standard SRGQMC trajectories, $H^{(1)}$ and $H^{(2)}$, are evolved independently, each using its own generator. Their difference isolates stochastic fluctuations and is used to estimate the source of the finite-walker bias. We use an auxiliary operator $B$ to track the accumulated bias, which gives the replica-corrected estimator
\begin{equation}
    H_{\mathrm{replica}}=\frac{H^{(1)}+H^{(2)}}{2}-B.
\end{equation}
In our SRGQMC(2) and SRGQMC(3) tests, this replica correction suppresses the finite-population bias below the statistical resolution. The auxiliary evolution equation, its derivation, and numerical validation are presented in the Supplemental Material~\cite{SM}.

\textit{Computational scaling and semi-stochastic evolution.\textemdash}
We show the practical advantage of SRGQMC using a detailed scaling analysis at $g=0.5\,\mathrm{MeV}$. Taking the normal-ordered three-body level as an example, we estimate the core evolution cost of SRGQMC(3) at a fixed $2\%$ correlation-energy accuracy. We find an effective scaling of $M^{3.4(2)}$ over single-particle basis sizes $M=8$--$16$~\cite{SM}, compared with the $M^9$ scaling of deterministic IMSRG(3). We then test a semi-stochastic strategy of SRGQMC(3), which deterministically evaluates commutator contributions with ranks no higher than two, while sampling the remaining higher-order contributions. This semi-stochastic method reduces statistical fluctuations and yields a further speedup of $1.5$--$2.2$, with a similar scaling of $M^{3.6(3)}$, demonstrating the promise of this hybrid approach. Stochastic sampling does not remove the algebraic complexity of higher-order commutators in the flow equation. Its practical cost depends on the sampling cost per walker and the total walker and loop numbers needed for a target accuracy, providing an adjustable tradeoff between precision and cost. Storing populated operator sites instead of full high-rank tensors can also reduce memory requirements.

SRGQMC is not free of sign cancellations. Positive and negative contributions to a commutator can be much larger than their net sum, increasing the sampling effort needed to resolve the flow. Annihilation combines opposite signs at each operator site but does not eliminate this statistical difficulty. For a given Hamiltonian, our semi-stochastic scheme performs the sign cancellations of selected low-order contractions deterministically, reducing sampling noise, which is analogous to explicit spin-isospin summation in GFMC~\cite{RevModPhys.87.1067}. This strategy thus mitigates cancellation-driven noise in SRGQMC calculations.

For many-body calculations in larger model spaces and with realistic nuclear interactions, large speedups are also possible if the stochastic sampling effort grows more slowly than the deterministic contraction cost. The hybrid semi-stochastic method offers a promising route to further improve the efficiency, although the computational cost for large-scale nuclear matter and finite nuclei calculations remains to be quantified.}

\end{document}


\title{Supplemental Material for: Stochastic Similarity Renormalization Group}

\author{R. Z. Hu\,\orcidlink{0009-0002-8797-6622}}
\thanks{These authors contributed equally to this work.}
\affiliation{School of Physics, and State Key Laboratory of Nuclear Physics and Technology, Peking University, Beijing 100871, China}
\author{X. Zhen\,\orcidlink{0009-0000-1806-4123}}
\thanks{These authors contributed equally to this work.}
\affiliation{School of Physics, and State Key Laboratory of Nuclear Physics and Technology, Peking University, Beijing 100871, China}
\author{F. R. Xu\,\orcidlink{0000-0001-6699-0965}}\email[Corresponding author: ]{frxu@pku.edu.cn}
\affiliation{School of Physics, and State Key Laboratory of Nuclear Physics and Technology, Peking University, Beijing 100871, China}
\affiliation{Southern Center for Nuclear-Science Theory (SCNT), Institute oxf Modern Physics, Chinese Academy of Sciences, Huizhou 516000, China}
\author{J. C. Pei\,\orcidlink{0000-0002-9286-1304}}\email[Corresponding author: ]{peij@pku.edu.cn}
\affiliation{School of Physics, and State Key Laboratory of Nuclear Physics and Technology, Peking University, Beijing 100871, China}
\affiliation{Southern Center for Nuclear-Science Theory (SCNT), Institute of Modern Physics, Chinese Academy of Sciences, Huizhou 516000, China}

\maketitle

\section{SRGQMC Algorithm Details}
In this section, we provide a more detailed description of the SRGQMC algorithm. Direct sampling in the full configuration space becomes prohibitive for large systems such as nuclear matter. We therefore place walkers on the matrix elements themselves, treating the evolved Hamiltonian as the sampled object in operator space. In this representation, the spawning and annihilation processes are designed to satisfy the SRG flow equations, yielding a stochastic realization of the evolution at each step. The commutator of normal-ordered operators generally induces operators of higher particle rank. For two normal-ordered operators containing up to \(m\)- and \(n\)-body terms, respectively, their commutator generates contributions ranging from \(|m-n|\)-body to \((m+n-1)\)-body operators,
\begin{align}
    \left[A^{(m)},B^{(n)}\right]
    =
    \sum_{k=|m-n|}^{m+n-1} C^{(k)} .
\end{align}
In IMSRG($n$), all operators and commutators are truncated at the normal-ordered $n$-body level. Once the relevant commutators have been evaluated, they can be used to construct the flow equations for both deterministic IMSRG and stochastic SRGQMC, differing only in whether the resulting evolution is represented and solved deterministically or stochastically.
It is worth noting that, unlike the situation encountered in most applications to finite nuclei, the momentum-space matrix elements in nuclear matter are generally complex-valued. As a result, several simplifications that rely on real-valued matrix elements are no longer applicable, and the resulting flow equations differ slightly from their more familiar forms used in finite nuclei calculations.

In the following expressions, the occupation number \(n_p\) is defined in the usual way as
\begin{align}
    n_p =
    \begin{cases}
    1, & \text{for hole states},\\
    0, & \text{for particle states}.
    \end{cases}
\end{align}
We also define $\bar{n}_p \equiv 1- n_p$. The exchange operators are defined by
\begin{align}
    P(pq\cdots/rs\cdots)
    \equiv
    1 - P_{pr} - P_{ps} - \cdots - P_{qr} - P_{qs} - \cdots ,
\end{align}
where the elementary exchange operator \(P_{ab}\) interchanges the indices \(a\) and \(b\) in the expression on which it acts. All matrix elements are assumed to be fully antisymmetrized.

A final algorithmic detail concerns the control of the total walker population. At $s=0$, the walker representation is normalized through
\begin{equation}
    \mathcal{N}(0) =
    \frac{1}{N_\mathrm{w}}\sum_{ij}|H_{ij}(0)|,
\end{equation}
so that $H_{ij}(0)=\mathcal{N}(0)C_{ij}(0)$ and the initial total population is fixed by the target value $N_\mathrm{w}$. During the subsequent evolution, spawning and annihilation generally change the instantaneous population $\widetilde N_\mathrm{w}=\sum_{ij}|\widetilde C_{ij}|$. After each step, we therefore apply a global rescaling $C_{ij}\leftarrow (N_\mathrm{w}/\widetilde N_\mathrm{w})\widetilde C_{ij}$ and update the physical scale factor as $\mathcal{N}\leftarrow \mathcal{N}\widetilde N_\mathrm{w}/N_\mathrm{w}$. The accumulated product of these scale factors is used when reconstructing the physical Hamiltonian and observables, so this population control changes only the stochastic representation, not the underlying SRGQMC estimate. \revise{Unlike population control through a fluctuating energy shift in FCIQMC~\cite{PhysRevB.103.155135}, the normalization used here is an exact rescaling of the walker representation. Every rescaling factor is retained in $\mathcal{N}$, so that the physical matrix elements $\mathcal{N}C_{ij}$ remain unchanged for each realization at every step. Ensemble averages are evaluated after reconstructing physical quantities, with all accumulated scale factors included. Consequently, this normalization introduces no additional population-control bias. The finite-walker bias addressed below instead originates from the nonlinear dependence of the SRG generator on the stochastically sampled Hamiltonian.}

\subsection{SRGQMC(1)}
There are two commutators included at this level:
\begin{align}
     	\comm{\AO^{(1)}}{\BO^{(1)}}^{(0)} &= \sum_{ij}A_{ij}B_{ji}(n_i-n_j),\\
  \comm{\AO^{(1)}}{\BO^{(1)}}^{(1)} &= \sum_{ij}\sum_a\nord{\aaO_i\aO_j}
 	\left(A_{ia}B_{aj}-B_{ia}A_{aj}\right).
\end{align}
The corresponding flow equations of IMSRG(1) are
\begin{align}
    \frac{{\rm d}E}{{\rm d}s} &= \sum_{ij}(n_i-n_j)\eta_{ij}f_{ji}, \\
    \frac{{\rm d}f_{ij}}{{\rm d}s} &= \sum_a\left(\eta_{ia}f_{aj} - f_{ia} \eta_{aj}\right).
\end{align}
These terms contribute for a general Hamiltonian. In nuclear matter, however, translational invariance ensures that the one-body operator \(f\) remains diagonal throughout the evolution. Consequently, the one-body part of the generator, \(\eta^{(1)}\), vanishes identically, and these terms do not contribute in the present case. We nevertheless include them here for completeness.

The situation is slightly different in SRGQMC. In this formulation, walkers reside on the matrix elements of \(H\) and stochastically spawn so as to reproduce the corresponding flow equations. Thus, the same flow equations should be interpreted in the opposite direction: instead of evaluating a given matrix element from all contributing terms in the deterministic sense, each occupied matrix element acts as a source that spawns walkers to the matrix elements to which it contributes. After relabeling the indices accordingly, the flow equations can be read directly as the spawning rules of walkers:
\begin{align}
    \Delta E &= -\Delta s\sum_{ab}(n_a-n_b) \eta_{ba}f_{ab},\label{eq:8} \\
    \Delta f_{ib} & = \Delta s \sum_a \eta_{ia}f_{ab},\label{eq:9} \\
    \Delta f_{ai} & = -\Delta s\sum_b \eta_{bi}f_{ab}\label{eq:10}.
\end{align}
We now use SRGQMC(1) as a simple example to illustrate the trajectory of a walker. Consider a walker located on the one-body matrix element \(f_{ab}\), which carries two indices \(a\) and \(b\). To satisfy Eq.~(\ref{eq:8}), this walker must identify the corresponding generator matrix element, \(\eta_{ba}=-\eta_{ab}^*\), and contribute to the increment of the normal-ordered zero-body energy through a simple product. For Eqs.~(\ref{eq:9}) and (\ref{eq:10}), however, the walker must keep one of its original indices fixed and randomly sample an additional index \(i\). Once \(i\) is selected, the walker contributes simultaneously to the matrix elements \(f_{ai}\) and \(f_{ib}\), with weights determined by the corresponding terms in the flow equation. Equivalently, the two spawning channels can be combined into a single heat-bath distribution. \revise{Let $w_{ab}$ denote the signed weight of an individual source walker on $f_{ab}$. The sum of all walker weights at site $(a,b)$ equals $C_{ab}$, and we recall that $C_{ab}\propto f_{ab}$. In this weighted-walker representation, these weights need not be integers. As an alternative to separate proposals for the two channels, the channel and target index can be sampled jointly from a heat-bath distribution.} Defining
\begin{align}
    Z_{ab}
    =
    \sum_i |\eta_{ia}|
    +
    \sum_i |\eta_{bi}|,
\end{align}
we choose the left channel \(f_{ab}\rightarrow f_{ib}\) with probability
\begin{align}
    P_L(i|a,b)
    =
    \frac{|\eta_{ia}|}{Z_{ab}},
\end{align}
and the right channel \(f_{ab}\rightarrow f_{ai}\) with probability
\begin{align}
    P_R(i|a,b)
    =
    \frac{|\eta_{bi}|}{Z_{ab}}.
\end{align}
The corresponding spawned weights are then
\begin{align}
    w_{ib}^{\mathrm{spawn}}
    &=
    \Delta s\,
    w_{ab}
    \frac{\eta_{ia}}{P_L(i|a,b)},
    \\
    w_{ai}^{\mathrm{spawn}}
    &=
    -\Delta s\,
    w_{ab}
    \frac{\eta_{bi}}{P_R(i|a,b)} .
\end{align}
\revise{Here \(P_L(i|a,b)\) and \(P_R(i|a,b)\) are joint probabilities for selecting both the channel and the target index, with \(\sum_i[P_L(i|a,b)+P_R(i|a,b)]=1\). Each draw selects one channel–target pair, and the corresponding signed weight is added to the target site. If \(Z_{ab}=0\), no spawning occurs through these channels.}
With this heat-bath sampling strategy, each individual term in the flow equations can be estimated at a given step. However, since the generator \(\eta\) depends on the stochastically sampled Hamiltonian \(H\), the full flow equation is globally nonlinear. In general,
\[
    \left\langle [\eta(H),H] \right\rangle
    \neq
    [\eta(\langle H\rangle),\langle H\rangle],
\]
so the stochastic evolution would accumulate a finite-population sampling bias relative to the exact deterministic flow.
\revise{Increasing the number of independent loops reduces the statistical uncertainty but does not remove finite-population bias. We address the latter through the replica method described below.}
After the spawning step, all walkers accumulated on $\Delta f$ are collected and annihilated (two walkers with different signs on the same site cancel with each other), yielding the total stochastic increment used for the next evolution step.

In the following discussion of SRGQMC(2) and SRGQMC(3), we no longer explain each term individually. Instead, we present the corresponding commutators and the flow equations in both the deterministic IMSRG and stochastic SRGQMC forms. Because the occupation-number structure becomes increasingly involved, we first introduce several compact notations. We define
\begin{align}
O^B_{ab} \equiv n_a - n_b,
\qquad
O^C_{ab} \equiv 1 - n_a - n_b ,
\end{align}
and also the symmetric and antisymmetric occupation-number combinations
\begin{align}
O^S_{ab\cdots/cd\cdots}
\equiv
n_a n_b \cdots \bar n_c \bar n_d \cdots
+
\bar n_a \bar n_b \cdots n_c n_d \cdots ,
\\
O^A_{ab\cdots/cd\cdots}
\equiv
n_a n_b \cdots \bar n_c \bar n_d \cdots
-
\bar n_a \bar n_b \cdots n_c n_d \cdots .
\end{align}
\subsection{SRGQMC(2)}
There are five more commutators included at this level:
\begin{align}
    \comm{\AO^{(1)}}{B^{(2)}}^{(1)} &= \sum_{ij}\sum_{ab}\nord{\aaO_i\aO_j}
 	O^B_{ab}A_{ab}B_{biaj}, \\
	\comm{\AO^{(1)}}{\BO^{(2)}}^{(2)} &= \frac{1}{4}\sum_{ijkl}\sum_{a}\nord{\aaO_i\aaO_j\aO_l\aO_k}
 	\left\{P(i/j) A_{ia}B_{ajkl}-P(k/l)A_{ak}B_{ijal} \right\}, \\
  \comm{\AO^{(2)}}{\BO^{(2)}}^{(0)} &= \frac{1}{4}\sum_{ijkl}O^A_{ij/kl}A_{ijkl}B_{klij}, \\
   	\comm{\AO^{(2)}}{\BO^{(2)}}^{(1)} &= \frac{1}{2}\sum_{ij}\sum_{abc} \nord{\aaO_i \aO_j}
 	O^S_{ab/c}\left(A_{ciab}B_{abcj}-B_{ciab}A_{abcj}\right),\\
  	\comm{\AO^{(2)}}{\BO^{(2)}}^{(2)} &= \frac{1}{4}\sum_{ijkl}\sum_{ab}\nord{\aaO_i\aaO_j\aO_l\aO_k}\biggl\{ \frac{1}{2}O^C_{ab}(A_{ijab}B_{abkl}-B_{ijab}A_{abkl})+O^B_{ab}P(i/j)P(k/l)A_{aibk}B_{bjal} \biggr\}.
\end{align}
The corresponding IMSRG(2) flow equations are
\begin{align}
    \frac{{\rm d}E}{{\rm d}s} & = \sum_{ij}O^B_{ij}\eta_{ij}f_{ji}+\frac{1}{4}\sum_{ijkl}O^A_{ij/kl}\eta_{ijkl}\Gamma_{klij}, \\
    \frac{{\rm d}f_{ij}}{{\rm d}s} &= \sum_a\left(\eta_{ia}f_{aj} - f_{ia} \eta_{aj}\right)
    + \sum_{ab}O^B_{ab}(\eta_{ab}\Gamma_{biaj}-f_{ab}\eta_{biaj}) + \frac{1}{2}\sum_{abc}
 	O^S_{ab/c}\left(\eta_{ciab}\Gamma_{abcj}-\Gamma_{ciab}\eta_{abcj}\right), \\
    \frac{{\rm d}\Gamma_{ijkl}}{{\rm d}s} &= \sum_a \left[P(i/j) \left(\eta_{ia}\Gamma_{ajkl} - f_{ia}\eta_{ajkl}\right) - P(k/l)\left(\eta_{ak}\Gamma_{ijal}-f_{ak}\eta_{ijal}\right)\right] \nonumber \\
    &+\sum_{ab}\biggl[ \frac{1}{2}O^C_{ab}(\eta_{ijab}\Gamma_{abkl}-\Gamma_{ijab}\eta_{abkl})+O^B_{ab}P(i/j)P(k/l)\eta_{aibk}\Gamma_{bjal} \biggr].
\end{align}
The standard flow equation above can also be found in Ref.~\cite{Hergert2016}.
The spawning rules of SRGQMC(2) are:
\begin{align}
    &\Delta E = -\Delta s\sum_{ab}O^B_{ab} \eta_{ba}f_{ab} + \frac{1}{4}\Delta s\sum_{abcd}O^A_{ab/cd}\eta_{abcd}\Gamma_{cdab}, \\
    &\Delta f_{ib}  = \Delta s \sum_a \eta_{ia}f_{ab},~\Delta f_{ai}  = -\Delta s\sum_b \eta_{bi}f_{ab}, \\
    &\Delta f_{ij}  = -\Delta s\sum_{ab} O^B_{ab}\eta_{biaj}f_{ab}, \\
    &\Delta \Gamma_{ajkl}, -\Delta \Gamma_{jakl} = -\Delta s \sum_{b} \eta_{bjkl}f_{ab},~\Delta \Gamma_{ijbl}, -\Delta \Gamma_{ijlb} = \Delta s \sum_{a} \eta_{ijal}f_{ab}, \\
    &\Delta f_{bd}  = \Delta s \sum_{ac}O^B_{ca}\eta_{ca}\Gamma_{abcd}, \\
    &\Delta \Gamma_{ibcd}, -\Delta \Gamma_{bicd} = \Delta s\sum_a \eta_{ia}\Gamma_{abcd},~\Delta \Gamma_{abid}, -\Delta \Gamma_{abdi} = -\Delta s\sum_{c} \eta_{ci}\Gamma_{abcd}, \\
    &\Delta f_{id} = \frac{1}{2}\Delta s \sum_{abc}O^S_{ab/c} \eta_{ciab}\Gamma_{abcd},~\Delta f_{bi} = -\frac{1}{2}\Delta s \sum_{acd} O^S_{cd/a} \eta_{cdai}\Gamma_{abcd}, \\
    &\Delta \Gamma_{ijcd} = \frac{1}{2}\Delta s \sum_{ab} O^C_{ab}\eta_{ijab}\Gamma_{abcd},~\Delta \Gamma_{abkl} = -\frac{1}{2}\Delta s \sum_{cd} O^C_{cd}\eta_{cdkl}\Gamma_{abcd}, \\
    &\Delta \Gamma_{ibjd}, -\Delta \Gamma_{bijd}, -\Delta \Gamma_{ibdj}, \Delta \Gamma_{bidj} = \Delta s\sum_{ac}O^B_{ca}\eta_{ciaj}\Gamma_{abcd}.
\end{align}
\subsection{SRGQMC(3)}
The IMSRG(3) flow equations have been presented in several previous works~\cite{Hergert2016,PhysRevC.103.044318,PhysRevC.110.044316}. Since these expressions are algebraically involved and minor typographical inconsistencies can easily arise, we have independently re-derived the following equations in order to ensure consistency and accuracy. 

There are ten more commutators appearing at this level:
\begin{align}
        \comm{\AO^{(1)}}{\BO^{(3)}}^{(2)} &= {\frac{1}{4}}\sum_{ijkl}\sum_{ab}\nord{\aaO_i\aaO_j\aO_l\aO_k}
     	O^B_{ab}A_{ab}B_{bijakl} \\
  \comm{\AO^{(1)}}{\BO^{(3)}}^{(3)} &= \frac{1}{36}\sum_{ijklmn}\sum_a\nord{\aaO_i\aaO_j\aaO_k \aO_n \aO_m \aO_l} \left\{P(i/jk)A_{ia}B_{ajklmn}-P(l/mn)A_{al}B_{ijkamn}  \right\} \\
  \comm{\AO^{(2)}}{\BO^{(2)}}^{(3)} &= \frac{1}{36}\sum_{ijklmn}\sum_a 
	\nord{\aaO_i\aaO_j\aaO_k \aO_n\aO_m\aO_l} \left\{P(ij/k)P(l/mn)\left(A_{ijla}B_{akmn}-B_{ijla}A_{akmn}\right)\right\} \\
 	\comm{\AO^{(2)}}{\BO^{(3)}}^{(1)} &= -\frac{1}{4}\sum_{ij}\sum_{abcd}\nord{\aaO_i\aO_j}
	O^A_{ab/cd}A_{cdab}B_{abijcd} \\
     \comm{\AO^{(2)}}{\BO^{(3)}}^{(2)} &= -\frac{1}{8}\sum_{ijkl}\sum_{abc}\nord{\aaO_i\aaO_j\aO_l\aO_k}O^S_{ab/c}\left\{P(i/j)A_{icab}B_{abjklc} - P(k/l)A_{abkc}B_{ijcabl}\right\} \\
 	\comm{\AO^{(2)}}{\BO^{(3)}}^{(3)} &= \frac{1}{36}\sum_{ijklmn}\sum_{ab}
 	\nord{\aaO_i\aaO_j\aaO_k\aO_n\aO_m\aO_l}\\
  &\times\Big\{\frac{1}{2}O^C_{ab}\left[P(ij/k)A_{ijab}B_{abklmn}-P(lm/n)A_{ablm}B_{ijkabn}\right]- O^B_{ab}P(ij/k)P(lm/n) A_{bkan}B_{ijalmb}\Big\}\nonumber \\
  \comm{\AO^{(3)}}{\BO^{(3)}}^{(0)} &= \frac{1}{36} \sum_{ijklmn}
	O^A_{ijk/lmn}
	A_{ijklmn}B_{lmnijk} \\
  	\comm{\AO^{(3)}}{\BO^{(3)}}^{(1)} &= \frac{1}{12} \sum_{ij}\sum_{abcde} \nord{\aaO_i\aO_j}  O^S_{ab/cde}
	(A_{abicde}B_{cdeabj}-B_{abicde}A_{cdeabj})\\
 	\comm{\AO^{(3)}}{\BO^{(3)}}^{(2)} &= \frac{1}{4}\sum_{ijkl}\sum_{abcd}
	\nord{\aaO_i\aaO_j\aO_l\aO_k}\nonumber\\
   &\times \Big\{\frac{1}{6}O^A_{a/bcd}
	(A_{aijbcd}B_{bcdakl}-A_{bcdakl}B_{aijbcd})-\frac{1}{4}O^A_{ab/cd}(1-P_{ij})(1-P_{kl})A_{abicdl}B_{cdjabk}
	\Big\} \\
 	\comm{\AO^{(3)}}{\BO^{(3)}}^{(3)}&= \frac{1}{36}\sum_{ijklmn}\sum_{abc}
	\nord{\aaO_i\aaO_j\aaO_k\aO_n\aO_m\aO_l}\\
  &\times\Big\{
	\frac{1}{6}O^S_{abc}
	(A_{ijkabc}B_{abclmn}-B_{ijkabc}A_{abclmn})-\frac{1}{2}O^S_{ab/c}P(ij/k){P(lm/n)}
	(A_{ijcabn}B_{abklmc}-A_{abklmc}B_{ijcabn})\nonumber
	\Big\}
\end{align}
The corresponding IMSRG(3) flow equations are
\begin{align}
    \frac{{\rm d}E}{{\rm d}s} & = \sum_{ij}O^B_{ij}\eta_{ij}f_{ji}+\frac{1}{4}\sum_{ijkl}O^A_{ij/kl}\eta_{ijkl}\Gamma_{klij}+\frac{1}{36} \sum_{ijklmn}
	O^A_{ijk/lmn}
	\eta_{ijklmn}W_{lmnijk}, \\
    \frac{{\rm d}f_{ij}}{{\rm d}s} &= \sum_a\left(\eta_{ia}f_{aj} - f_{ia} \eta_{aj}\right)
    + \sum_{ab}O^B_{ab}(\eta_{ab}\Gamma_{biaj}-f_{ab}\eta_{biaj}) + \frac{1}{2}\sum_{abc}
 	O^S_{ab/c}\left(\eta_{ciab}\Gamma_{abcj}-\Gamma_{ciab}\eta_{abcj}\right)\nonumber\\
    &-\frac{1}{4}\sum_{abcd}
	O^A_{ab/cd}(\eta_{cdab}W_{abijcd}-\Gamma_{cdab}\eta_{abijcd})+\frac{1}{12}\sum_{abcde}O^S_{ab/cde}
	(\eta_{abicde}W_{cdeabj}-W_{abicde}\eta_{cdeabj}), \\
    \frac{{\rm d}\Gamma_{ijkl}}{{\rm d}s} &= \sum_a \left[P(i/j) \left(\eta_{ia}\Gamma_{ajkl} - f_{ia}\eta_{ajkl}\right) - P(k/l)\left(\eta_{ak}\Gamma_{ijal}-f_{ak}\eta_{ijal}\right)\right] +\sum_{ab}O^B_{ab}(\eta_{ab}W_{bijakl}-f_{ab}\eta_{bijakl})\nonumber \\
    &+\sum_{ab}\biggl[ \frac{1}{2}O^C_{ab}(\eta_{ijab}\Gamma_{abkl}-\Gamma_{ijab}\eta_{abkl})+O^B_{ab}P(i/j)P(k/l)\eta_{aibk}\Gamma_{bjal} \biggr]\nonumber\\
    & -\frac{1}{2}\sum_{abc}O^S_{ab/c}\left\{P(i/j)(\eta_{icab}W_{abjklc}-\Gamma_{icab}\eta_{abjklc}) - P(k/l)(\eta_{abkc}W_{ijcabl}-\Gamma_{abkc}\eta_{ijcabl})\right\} \nonumber\\
    &+\sum_{abcd}\Big\{\frac{1}{6}O^A_{a/bcd}
	(\eta_{aijbcd}W_{bcdakl}-W_{aijbcd}\eta_{bcdakl})-\frac{1}{4}O^A_{ab/cd}(1-P_{ij})(1-P_{kl})\eta_{abicdl}W_{cdjabk}
	\Big\},\\
    \frac{{\rm d}W_{ijklmn}}{{\rm d}s} &= \sum_a\left\{P(i/jk)(\eta_{ia}W_{ajklmn}-f_{ia}\eta_{ajklmn})-P(l/mn)(\eta_{al}W_{ijkamn}-f_{al}\eta_{ijkamn})  \right\}\\
    & + \sum_a \left\{P(ij/k)P(l/mn)\left(\eta_{ijla}\Gamma_{akmn}-\Gamma_{ijla}\eta_{akmn}\right)\right\} -\sum_{ab} O^B_{ab}P(ij/k)P(lm/n) (\eta_{bkan}W_{ijalmb}-\Gamma_{bkan}\eta_{ijalmb})\nonumber\\
    &+\sum_{ab}\Big\{\frac{1}{2}O^C_{ab}\left[P(ij/k)(\eta_{ijab}W_{abklmn}-\Gamma_{ijab}\eta_{abklmn})-P(lm/n)(\eta_{ablm}W_{ijkabn}-\Gamma_{ablm}\eta_{ijkabn})\right]\Big\}\nonumber \\
    &+\sum_{abc}\Big\{
	\frac{1}{6}O^S_{abc}
	(\eta_{ijkabc}W_{abclmn}-W_{ijkabc}\eta_{abclmn})-\frac{1}{2}O^S_{ab/c}P(ij/k){P(lm/n)}
	(\eta_{ijcabn}W_{abklmc}-W_{ijcabn}\eta_{abklmc})\nonumber
	\Big\}.
\end{align}
The spawning rules of SRGQMC(3) are:
\begin{align}
    &\Delta E = -\Delta s\sum_{ab}O^B_{ab} \eta_{ba}f_{ab} + \frac{1}{4}\Delta s\sum_{abcd}O^A_{ab/cd}\eta_{abcd}\Gamma_{cdab} + \frac{1}{36}\Delta s \sum_{abcdef}O^A_{abc/def}\eta_{abcdef}W_{defabc}, \\
    &\Delta f_{ib}  = \Delta s \sum_a \eta_{ia}f_{ab},~ \Delta f_{ai}  = -\Delta s\sum_b \eta_{bi}f_{ab}, \\
    &\Delta f_{ij}  = -\Delta s\sum_{ab} O^B_{ab}\eta_{biaj}f_{ab}, \\
    &\Delta \Gamma_{ajkl}, -\Delta \Gamma_{jakl} = -\Delta s \sum_{b} \eta_{bjkl}f_{ab},~ \Delta \Gamma_{ijbl}, -\Delta \Gamma_{ijlb} = \Delta s \sum_{a} \eta_{ijal}f_{ab}, \\
    &\Delta \Gamma_{ijkl} = \Delta s\sum_{ab} O^B_{ba}\eta_{bijakl}f_{ab}, \\
    &\Delta W_{ajklmn}, -\Delta W_{jaklmn}, -\Delta W_{kjalmn} = -\Delta s\sum_{b}\eta_{bjklmn}f_{ab}, \\
    &\Delta W_{ijkbmn}, -\Delta W_{ijkmbn}, -\Delta W_{ijknmb} = \Delta s \sum_{a}\eta_{ijkamn}f_{ab},\\
    &\Delta f_{bd}  = \Delta s \sum_{ac}O^B_{ca}\eta_{ca}\Gamma_{abcd}, \\
    &\Delta \Gamma_{ibcd}, -\Delta \Gamma_{bicd} = \Delta s\sum_a \eta_{ia}\Gamma_{abcd}, ~\Delta \Gamma_{abid}, -\Delta \Gamma_{abdi} = -\Delta s\sum_{c} \eta_{ci}\Gamma_{abcd}, \\
    &\Delta f_{id} = \frac{1}{2}\Delta s \sum_{abc}O^S_{ab/c} \eta_{ciab}\Gamma_{abcd},~\Delta f_{bi} = -\frac{1}{2}\Delta s \sum_{acd} O^S_{cd/a} \eta_{cdai}\Gamma_{abcd}, \\
    &\Delta \Gamma_{ijcd} = \frac{1}{2}\Delta s \sum_{ab} O^C_{ab}\eta_{ijab}\Gamma_{abcd}, ~\Delta \Gamma_{abkl} = -\frac{1}{2}\Delta s \sum_{cd} O^C_{cd}\eta_{cdkl}\Gamma_{abcd}, \\
    &\Delta \Gamma_{ibjd}, -\Delta \Gamma_{bijd}, -\Delta \Gamma_{ibdj}, \Delta \Gamma_{bidj} = \Delta s\sum_{ac}O^B_{ca}\eta_{ciaj}\Gamma_{abcd}, \\
    & P(ij/b)P(l/cd)\Delta W_{ijblcd} = \Delta s \sum_{a} \eta_{ijla}\Gamma_{abcd},~ P(ab/k)P(c/mn)\Delta W_{abkcmn} = - \Delta s \sum_d \eta_{dkmn}\Gamma_{abcd}, \\
    & \Delta f_{ij} = \frac{1}{4}\Delta s \sum_{abcd} O^A_{cd/ab}\eta_{cdijab}\Gamma_{abcd}, \\
    & \Delta \Gamma_{ajkl}, -\Delta \Gamma_{jakl} = \frac{1}{2}\Delta s \sum_{bcd}O^S_{cd/b}\eta_{cdjklb}\Gamma_{abcd}, ~\Delta \Gamma_{ijcl}, -\Delta \Gamma_{ijlc} = -\frac{1}{2}\Delta s \sum_{abd}O^S_{ab/d}\eta_{ijdabl}\Gamma_{abcd}, \\
    & P(ij/b)P(lm/d)\Delta W_{ijblmd} = \Delta s\sum_{ac} O^B_{ca}\eta_{ijclma}\Gamma_{abcd}, \\
    & P(ab/k) \Delta W_{abklmn} = -\frac{1}{2}\Delta s\sum_{cd} O^C_{cd}\eta_{cdklmn}\Gamma_{abcd}, ~P(cd/n) \Delta W_{ijkcdn} = \frac{1}{2}\Delta s\sum_{ab}O_{ab}^C\eta_{ijkabn}\Gamma_{abcd},\\
    &\Delta \Gamma_{bcef} = \Delta s \sum_{ad}O^B_{da}\eta_{da}W_{abcdef}, \\
    & P(i/bc)\Delta W_{ibcdef} = \Delta s \sum_{a}\eta_{ia}W_{abcdef},~ P(l/ef)\Delta W_{abclef} = -\Delta s \sum_d \eta_{dl}W_{abcdef}, \\
    & \Delta f_{cd} = -\frac{1}{4}\Delta s \sum_{abef}O^A_{ab/ef}\eta_{efab}W_{abcdef}, \\
    & P(i/c) \Delta \Gamma_{icde} = -\frac{1}{2}\Delta s \sum_{abf} O^S_{ab/f}\eta_{ifab}W_{abcdef}, ~P(k/f) \Delta \Gamma_{abkf} = \frac{1}{2}\Delta s \sum_{cde}O^S_{de/c}\eta_{dekc}W_{abcdef}, \\
    & P(ij/c)\Delta W_{ijcdef} = \frac{1}{2}\Delta s \sum_{ab}O^C_{ab}\eta_{ijab}W_{abcdef},~ P(lm/f)\Delta W_{abclmf} = -\frac{1}{2}\Delta s \sum_{de}O^C_{de}\eta_{delm}W_{abcdef},\\
    & P(ab/k)P(de/n)\Delta W_{abkden} = -\Delta s \sum_{cf}O_{cf}^B\eta_{fkcn}W_{abcdef}, \\
    & \Delta f_{if} = \frac{1}{12} \Delta s \sum_{abcde} O_{de/abc}^S \eta_{deiabc}W_{abcdef}, ~ \Delta f_{cj} = -\frac{1}{12} \Delta s \sum_{abdef} O_{ab/def}^S\eta_{defabj}W_{abcdef}, \\
    & \Delta \Gamma_{ijef} = \frac{1}{6}\Delta s \sum_{abcd}O_{d/abc}^A\eta_{dijabc}W_{abcdef},~ \Delta \Gamma_{bckl} = -\frac{1}{6}\Delta s \sum_{adef}O_{a/def}^A \eta_{defakl}W_{abcdef}, \\
    & P(i/c)P(f/l)\Delta\Gamma_{icfl} = \frac{1}{4}\Delta s\sum_{abde}O_{ab/de}^A\eta_{deiabl}W_{abcdef}, \\
    & \Delta W_{ijkdef} = \frac{1}{6} \Delta s \sum_{abc}O^S_{abc}\eta_{ijkabc}W_{abcdef},~\Delta W_{abclmn} =-\frac{1}{6} \Delta s \sum_{def}O^S_{def}\eta_{deflmn}W_{abcdef},\\
    &P(ij/c)P(de/n)\Delta W_{ijcden} =-\frac{1}{2}\Delta s \sum_{abf} O^S_{ab/f}\eta_{ijfabn}W_{abcdef}, \\
    & P(ab/k)P(lm/f)\Delta W_{abklmf} =\frac{1}{2}\Delta s\sum_{cde}O_{de/c}^S\eta_{deklmc}W_{abcdef}.
\end{align}

\section{IMSRG(4) Flow Equation}
To the best of our knowledge, no full implementation (even the formalism) of the IMSRG truncated at the normal-ordered four-body level [IMSRG(4)] has been reported to date. Here, we present the IMSRG(4) flow equations, from which the corresponding SRGQMC(4) formalism was derived using a procedure similar to that in the derivations of SRGQMC(2)/(3). While the tedious formulas are omitted here, the SRGQMC(4) framework follows those of SRGQMC(2)/(3) but explicitly includes four-body contributions in the Hamiltonian and other operators. It is worth noting that recent automated diagram-generation tools~\cite{chen2026,Tichai2022} provide valuable means of cross-checking lengthy many-body commutator formulas. All matrix elements are assumed to be fully antisymmetrized. For compactness, we denote an $n$-body operator $O^{(n)}$ simply by $n$, and suppress the explicit normal-ordered strings in the resulting commutators. No Hermiticity assumption is imposed on the operators. At the IMSRG(4) level, the commutator $[A,B]$ contains the following 16 additional terms beyond IMSRG(3).

\subsection{$[1,\circ]$}
\begin{align}
    [1,4]\rightarrow 3 & = \frac{1}{36}\sum_{ijklmn}\sum_{ab}\left\{(n_a-n_b)A_{ab}B_{ijkblmna}\right\}, \\
  [1,4]\rightarrow 4 & = \frac{1}{576}\sum_{ijklmnpq}\sum_{a}\left\{P(i/jkl)A_{ia}B_{ajklmnpq}-P(m/npq)A_{am}B_{ijklanpq}\right\}.
\end{align}

\subsection{$[2,\circ]$}
\begin{align}
    [2,3]\rightarrow 4 & = \frac{1}{576}\sum_{ijklmnpq}\sum_{a}\Big\{(1 - P_{ik} -P_{il} - P_{jk} - P_{jl} + P_{ik}P_{jl})P(m/npq)A_{ijma}B_{klanpq}\nonumber\\
  & - (1-P_{mp}-P_{mq}-P_{np}-P_{nq}+P_{mp}P_{nq})P(i/jkl)A_{iamn}B_{jklpqa}\Big\},\\
    [2,4]\rightarrow 2 &= \frac{1}{16}\sum_{ijkl}\sum_{abcd}(\bar{n}_a\bar{n}_bn_cn_d-n_an_b \bar{n}_c\bar{n}_d)A_{cdab}B_{abijklcd}, \\
    [2,4]\rightarrow 3 &= \frac{1}{72}\sum_{ijklmn}\sum_{abc}
 	(n_an_b\bar{n}_c+\bar{n}_a\bar{n}_bn_c)\left\{P(i/jk)A_{icab}B_{jkablmnc} - P(l/mn)A_{ablc}B_{ijkcmnab}\right\}, \\
  [2,4] \rightarrow 4 & = \frac{1}{576}\sum_{ijklmnpq}\sum_{ab}\Bigg\{\frac{1}{2}(1-n_a-n_b)\Big[(1 - P_{ik} -P_{il} - P_{jk} - P_{jl} + P_{ik}P_{jl})A_{ijab}B_{abklmnpq} \\
  & - (1-P_{mp}-P_{mq}-P_{np}-P_{nq}+P_{mp}P_{nq})A_{abmn}B_{ijklpqab}\Big]+ (n_a-n_b)P(i/jkl)P(m/npq)A_{iamb}B_{jklbnpqa}\Bigg\}\nonumber.
\end{align}

\subsection{$[3,\circ]$}
\begin{align}
    [3,3]\rightarrow 4&= \frac{1}{576}\sum_{ijklmnpq}\sum_{ab}\Bigg\{\frac{1}{2}(1-n_a-n_b)\Big[P(ijk/l)P(m/npq)A_{ijkmab}B_{ablnpq}- P(i/jkl)P(mnp/q)A_{abimnp}B_{jklqab}\Big] \nonumber\\
  &-(n_a-n_b)(1 - P_{ik} -P_{il} - P_{jk} - P_{jl} + P_{ik}P_{jl})(1-P_{mp}-P_{mq}-P_{np}-P_{nq}+P_{mp}P_{nq})A_{ijbmna}B_{klapqb}\Bigg\},\\
    [3,4] \rightarrow 1 &= \frac{1}{36} \sum_{ij}\sum_{abcdef}
    (n_an_bn_c\bar{n}_d\bar{n}_e\bar{n}_f - \bar{n}_a\bar{n}_b\bar{n}_cn_dn_en_f)A_{abcdef}B_{idefjabc},
    \\
    [3,4] \rightarrow 2 &= -\frac{1}{48}\sum_{ijkl}\sum_{abcde}
	(n_an_bn_c\bar{n}_d\bar{n}_e + \bar{n}_a\bar{n}_b\bar{n}_cn_dn_e)\Big[P(i/j)A_{ideabc}B_{jabcklde} - P(k/l)A_{abckde}B_{ijdelabc}\Big], \\
 [3,4]\rightarrow 3 &= \frac{1}{36}\sum_{ijklmn}\sum_{abcd}\Bigg\{\frac{1}{6}(n_a\bar{n}_b\bar{n}_c\bar{n}_d-\bar{n}_a n_bn_cn_d) 
	\Big[P(ij/k)A_{aijbcd}B_{kbcdlmna}-P(lm/n)A_{bcdalm}B_{ijkanbcd}\Big] \nonumber\\ &+\frac{1}{4}(\bar{n}_a\bar{n}_b n_c n_d-n_an_b\bar{n}_c\bar{n}_d) P(i/jk)P(l/mn)A_{cdiabl}B_{abjkcdmn}
	\Bigg\},\\
 [3,4]\rightarrow 4&= \frac{1}{576}\sum_{ijklmnpq}\sum_{abc}\Bigg\{\frac{1}{6}(n_an_bn_c + \bar{n}_a\bar{n}_b\bar{n}_c)\Big[P(ijk/l)A_{ijkabc}B_{abclmnpq} - P(mnp/q)A_{abcmnp}B_{ijklabcq}\Big] \nonumber\\
 &+ \frac{1}{2} (n_a\bar{n}_b\bar{n}_c + \bar{n}_an_bn_c)\Big[(1 - P_{ik} -P_{il} - P_{jk} - P_{jl} + P_{ik}P_{jl})P(m/npq)A_{ijambc}B_{klbcnpqa} \nonumber\\
 &- (1-P_{mp}-P_{mq}-P_{np}-P_{nq}+P_{mp}P_{nq})P(i/jkl)A_{ibcmna}B_{jklapqbc}\Big]\Bigg\}.
\end{align}

\subsection{$[4,\circ]$}
\begin{align}
     [4,4]\rightarrow 0  &= \frac{1}{576} \sum_{ijklabcd}
	(n_in_jn_kn_l\bar{n}_a\bar{n}_b\bar{n}_c\bar{n}_d-\bar{n}_i\bar{n}_j\bar{n}_k \bar{n}_l n_a n_b n_c n_d)
	A_{ijklabcd}B_{abcdijkl},\\
[4,4] \rightarrow 1 &= \frac{1}{144} \sum_{ij}\sum_{abcdefg} (n_an_bn_cn_d\bar{n}_e\bar{n}_f\bar{n}_g + \bar{n}_a\bar{n}_b\bar{n}_c\bar{n}_dn_en_fn_g)(A_{iefgabcd}B_{abcdjefg} - A_{abcdjefg}B_{iefgabcd}), \\
[4,4] \rightarrow 2 &= \frac{1}{4}\sum_{ijkl}\sum_{abcdef}
	\Bigg\{\frac{1}{48}(n_an_b\bar{n}_c\bar{n}_d\bar{n}_e\bar{n}_f - \bar{n}_a\bar{n}_bn_cn_dn_en_f)(A_{abijcdef}B_{cdefklab} - A_{cdefklab}B_{abijcdef})\nonumber\\
 &+ \frac{1}{36} (n_an_bn_c\bar{n}_d\bar{n}_e\bar{n}_f-\bar{n}_a\bar{n}_b\bar{n}_cn_dn_en_f)(1-P_{ij})(1-P_{kl})A_{iabckdef}B_{jdeflabc}\Bigg\} ,\\
 [4,4] \rightarrow 3 &= \frac{1}{36}\sum_{ijklmn}\sum_{abcde} \Bigg\{\frac{1}{24}(n_an_bn_cn_d\bar{n}_e + \bar{n}_a\bar{n}_b\bar{n}_c\bar{n}_dn_e)(A_{ijkedabc}B_{dabclmne} - A_{cdablmne}B_{ijkecdab})\nonumber\\
 + \frac{1}{12}(n_a&n_bn_c\bar{n}_d\bar{n}_e + \bar{n}_a\bar{n}_b\bar{n}_cn_dn_e)\left[P(i/jk)P(lm/n)A_{iabclmde}B_{jkdenabc} - P(ij/k)P(l/mn)A_{ijdelabc}B_{kabcmnde}\right]\Bigg\}, \\
[4,4] \rightarrow 4 &= \frac{1}{576}\sum_{ijklmnpq}\sum_{abcd} \Bigg\{ \frac{1}{24}(\bar{n}_a\bar{n}_b\bar{n}_c\bar{n}_d - n_an_bn_cn_d) (A_{ijklabcd}B_{abcdmnpq} - B_{ijklabcd}A_{abcdmnpq})\nonumber\\
&+ \frac{1}{6}(n_a\bar{n}_b\bar{n}_c\bar{n}_d - \bar{n}_an_bn_cn_d)\Big[P(ijk/l)P(m/npq)A_{ijkambcd}B_{lbcdnpqa}- P(i/jkl)P(mnp/q)A_{ibcdmnpa}B_{jklaqbcd}\Big]\nonumber \\
+ &\frac{1}{4}(n_an_b\bar{n}_c\bar{n}_d - \bar{n}_a\bar{n}_bn_cn_d)\times(1 - P_{ik} -P_{il} - P_{jk} - P_{jl} + P_{ik}P_{jl})\nonumber\\
&(1-P_{mp}-P_{mq}-P_{np}-P_{nq}+P_{mp}P_{nq})A_{ijabmncd}B_{klcdpqab}\Bigg\}.
\end{align}
While a deterministic IMSRG(4) calculation remains unimplemented due to prohibitive computational costs, the SRGQMC(4) delivers an IMSRG(4) computation by using the stochastic quantum Monte Carlo technique. 

\section{Uncertainty Analysis}

\revise{In this section we examine statistical convergence and finite-population bias in SRGQMC. We first discuss statistical uncertainties in the standard stochastic evolution, and then introduce the replica method which can significantly suppress finite-population bias.}

\begin{figure*}[t]
    \centering
    \includegraphics[width=0.95\textwidth]{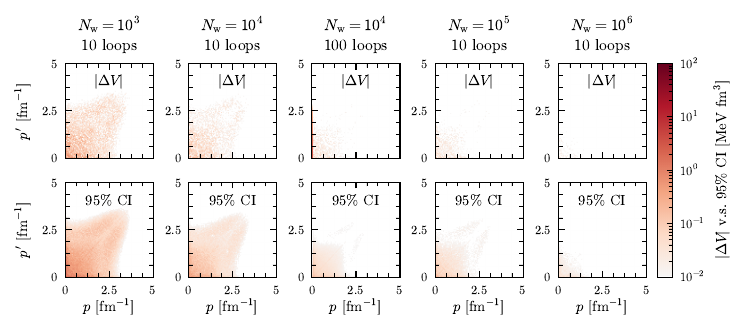}
    \caption{Convergence of the SRGQMC-evolved N$^3$LO$_{\rm EMN}(500)$ NN interaction in the $^1S_0$ channel at $\lambda=2.0\,\mathrm{fm}^{-1}$. From left to right, $(N_\mathrm{w},N_{\rm loop})=(10^3,10)$, $(10^4,10)$, $(10^4,100)$, $(10^5,10)$, and $(10^6,10)$. The upper row shows the absolute difference $|\Delta V|=|\bar V-V_{\rm SRG}|$ between the loop-averaged SRGQMC result and the deterministic SRG result. The lower row shows the half-width of the pointwise 95\% confidence intervals of the SRGQMC results.}
    \label{fig:nn_convergence}
\end{figure*}

\subsection{Statistical uncertainty with $N_\mathrm{w}$- and $N_\mathrm{loop}$-convergence}
For an observable $O$ (such as energy or phase shift), each independent stochastic evolution gives one estimate $O^{(\alpha)}$. The quoted central value and statistical uncertainty are taken as
\begin{align}
    \bar O =
    \frac{1}{N_{\rm loop}}\sum_{\alpha=1}^{N_{\rm loop}} O^{(\alpha)},\qquad
    \Delta O =
    \sqrt{\frac{1}{N_{\rm loop}(N_{\rm loop}-1)}
    \sum_{\alpha=1}^{N_{\rm loop}}\left(O^{(\alpha)}-\bar O\right)^2}.
\end{align}
Here $\Delta O$ is the standard error of the mean over independent loops; it does not include systematic bias. Increasing $N_{\rm loop}$ reduces the statistical error of the final averaged estimator, while increasing the walker population $N_\mathrm{w}$ improves the stochastic resolution of each individual SRGQMC trajectory and can reduce finite-population effects.

\revise{To quantify the overall agreement of the evolved interaction matrix
elements with the deterministic SRG results, we define the relative
root-mean-square (RMS) deviation as
\begin{equation}
    \epsilon_{\mathrm{RMS}}
    =
    \sqrt{
    \frac{\sum_{ij}|\bar V_{ij}-V_{\mathrm{SRG},ij}|^2}
         {\sum_{ij}|V_{\mathrm{SRG},ij}|^2}}
    ,
    \label{eq:relative_rms}
\end{equation}
where $\bar V$ is the loop-averaged SRGQMC interaction and
$V_{\mathrm{SRG}}$ is the deterministic reference.
We report this dimensionless deviation as a percentage.}

Figure~\ref{fig:nn_convergence} examines the convergence directly at the level of the evolved NN interaction matrix elements. Increasing $N_\mathrm{w}$ at fixed $N_{\rm loop}=10$ reduces the fluctuations and the estimated uncertainties. Increasing $N_{\rm loop}$ from 10 to 100 at fixed $N_\mathrm{w}=10^4$ further reduces the statistical uncertainties, as shown by the narrower confidence intervals in the lower row.

\begin{figure*}[t]
    \centering
    \includegraphics[width=0.95\textwidth]{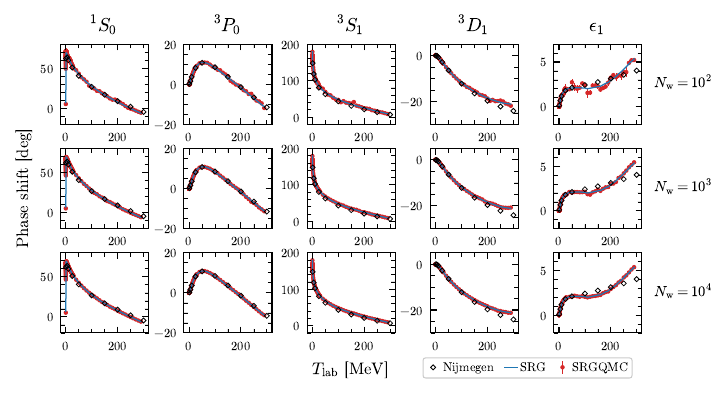}
    \caption{Convergence of the calculated phase shifts with respect to the walker population $N_\mathrm{w}$. Three rows correspond to $N_\mathrm{w}=10^2$, $10^3$, and $10^4$, respectively, with $N_{\rm loop}=10$ fixed. Columns show the $^1S_0$, $^3P_0$, $^3S_1$, $^3D_1$, and $\epsilon_1$ channels for the N$^3$LO$_{\rm EMN}(500)$ interaction evolved to $\lambda=2.0\,\mathrm{fm}^{-1}$. Solid curves denote deterministic SRG results, red points with error bars denote SRGQMC results, and open diamonds show the Nijmegen phase-shift analysis as an empirical reference.}
    \label{fig:phase_convergence}
\end{figure*}

\begin{figure*}[t]
    \centering
    \includegraphics[width=0.5\textwidth]{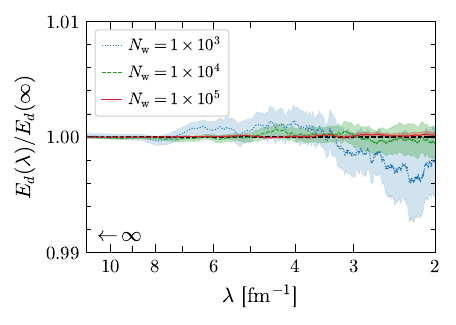}
    \caption{Evolution of deuteron binding energy $E_d$ during the SRGQMC evolution, where $E_d(\lambda)$ denotes the averaged $E_d$ calculated from the SRGQMC-evolved interaction at resolution scale $\lambda$. $E_d(\infty)$ denotes the deuteron binding energy of the bare interaction, which should remain unchanged if the unitarity of the SRG transformation is accurately preserved. $N_\mathrm{loop}=10$ is used.}
    \label{fig:deuteron_convergence}
\end{figure*}

Figure~\ref{fig:phase_convergence} examines the $N_\mathrm{w}$ convergence of the free-space NN calculation through scattering phase shifts. This is a stringent test because the phase shifts are not sampled directly; they are obtained after solving the NN scattering problem with the stochastically evolved interaction. Already for $N_\mathrm{w}=10^2$, the SRGQMC phase shifts follow the deterministic SRG curves in all partial waves, with rather large statistical uncertainties in some channels. Increasing the population to $N_\mathrm{w}=10^3$ and then to $10^4$ systematically reduces the statistical error bars and generally improves the agreement with the deterministic SRG results.
\revise{We also test the $N_\mathrm{w}$ convergence for the deuteron binding energy during the SRGQMC evolution, which tests the unitarity of the SRG transformation in our stochastic formalism. As shown in Fig.~\ref{fig:deuteron_convergence}, when using $N_\mathrm{w}=10^3$, the central value of deuteron binding energy $E_d(\lambda)$ begins to deviate from $E_d(\infty)$ when $\lambda \lesssim 4.0$ fm$^{-1}$, although the largest relative deviation is only about 0.4\%. When increasing $N_\mathrm{w}$ to $10^4$ and $10^5$, both this deviation and the statistical uncertainty are systematically reduced. These tests show that increasing $N_\mathrm{w}$ is an efficient way to reduce statistical and systematic uncertainty in SRGQMC calculations.}

\begin{figure*}[t]
    \centering
    \includegraphics[width=0.95\textwidth]{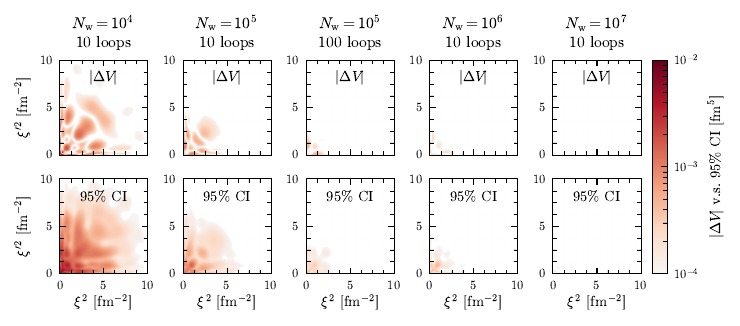}
    \caption{Convergence of the free-space SRGQMC evolution of the 3N interaction at $\lambda=2.0\,\mathrm{fm}^{-1}$. The interaction and displayed channel are the same as in the End Matter of the main text: $\mathcal J^\pi=\tfrac12^+$, $\mathcal T=\tfrac12$, $\alpha'=\alpha=0$, and $\tan\theta=\tan\theta'=\pi/4$. The hypermomenta $\xi,\xi'$ and hyperangles $\theta,\theta'$ are also defined in the End Matter. From left to right, we show results with $(N_\mathrm{w},N_{\rm loop})=(10^4,10)$, $(10^5,10)$, $(10^5,100)$, $(10^6,10)$, and $(10^7,10)$. As in Fig.~\ref{fig:nn_convergence}, the upper and lower rows show $|\Delta V|$ and the pointwise 95\% confidence-interval half-width, respectively, with a common logarithmic color scale in $\mathrm{fm}^5$.}
    \label{fig:3n_convergence}
\end{figure*}

\revise{Figure~\ref{fig:3n_convergence} extends these convergence tests to the free-space 3N interaction evolution. At fixed $N_{\rm loop}=10$, increasing $N_\mathrm{w}$ from $10^4$ to $10^7$ reduces both the fluctuations and the statistical uncertainties. The relative RMS deviation from the deterministic SRG result in the displayed channel decreases from about $3.95\%$ to $0.14\%$. At fixed $N_\mathrm{w}=10^5$, increasing $N_{\rm loop}$ from 10 to 100 narrows the confidence intervals and reduces the relative RMS deviation from $1.54\%$ to $0.58\%$. These results show that increasing either the walker population or the number of independent loops improves the statistical precision of the sampled 3N interaction and its agreement with the deterministic evolution.}

\begin{figure*}[t]
    \centering
    \includegraphics[width=0.98\textwidth]{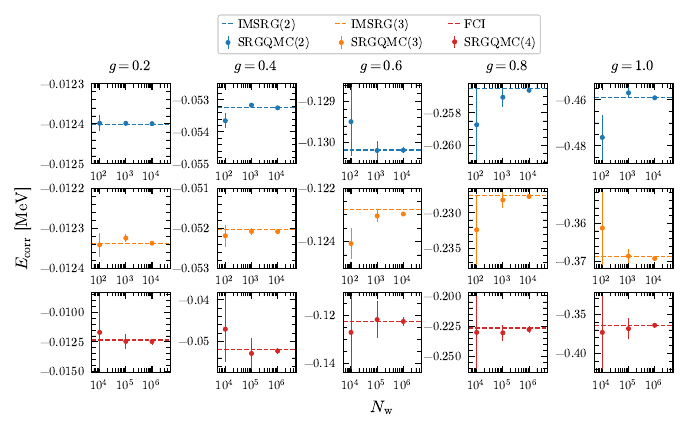}
    \caption{Convergence of the Richardson pairing model correlation energy with respect to $N_\mathrm{w}$. Columns correspond to $g=0.2$, $0.4$, $0.6$, $0.8$, and $1.0\,\mathrm{MeV}$. The three rows show SRGQMC(2), SRGQMC(3), and SRGQMC(4), respectively. Dashed horizontal lines denote the corresponding deterministic IMSRG(2), IMSRG(3) and FCI values.}
    \label{fig:pairing_nw_convergence}
\end{figure*}

Similar behavior is found in in-medium many-body calculations shown in Fig.~\ref{fig:pairing_nw_convergence}. We consider the Richardson pairing model at five representative coupling strengths, $g=0.2$, $0.4$, $0.6$, $0.8$, and $1.0\,\mathrm{MeV}$, and compare SRGQMC results with their deterministic counterparts. For SRGQMC(2) and SRGQMC(3), walker populations $N_\mathrm{w}=10^2$, $10^3$, and $10^4$ are compared with deterministic IMSRG(2) and IMSRG(3), respectively. For SRGQMC(4), the larger operator space requires larger populations, and we show $N_\mathrm{w}=10^4$, $10^5$, and $10^6$, compared with FCI. \revise{The agreement of SRGQMC(4) and FCI results does not imply formal exactness of IMSRG(4)
for $A=4$, since recovering the FCI limit generally requires retaining
normal-ordered operators up to rank $2A$~\cite{PhysRevC.110.044316}, corresponding here to
IMSRG(8).}

Across the coupling strengths and truncation orders shown, the SRGQMC estimates agree better with the corresponding reference values when increasing the total walker number $N_\mathrm{w}$. As $N_\mathrm{w}$ is increased, the statistical error decreases and the stochastic estimates approach the reference lines. This trend persists from the weak-coupling regime to the strongest coupling shown.

\begin{figure*}[t]
    \centering
    \includegraphics[width=0.98\textwidth]{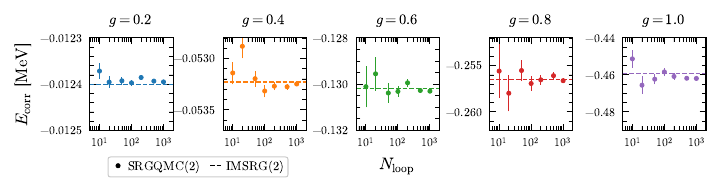}
    \caption{Convergence of SRGQMC(2) correlation energies with respect to the number of independent loops $N_{\rm loop}$ at fixed walker population $N_\mathrm{w}=100$. Panels correspond to $g=0.2$, $0.4$, $0.6$, $0.8$, and $1.0\,\mathrm{MeV}$. Points with error bars are SRGQMC(2) estimates obtained from independent-loop samples, while dashed horizontal lines denote deterministic IMSRG(2) results.}
    \label{fig:pairing_loop_convergence}
\end{figure*}

Figure~\ref{fig:pairing_loop_convergence} shows the role of the number of independent stochastic loops $N_\mathrm{loop}$. Here we fix the walker population to the deliberately small value $N_\mathrm{w}=100$ and focus on SRGQMC(2), comparing with deterministic IMSRG(2) at the same five pairing strengths. The loop number is varied from $N_{\rm loop}=10$ to $1000$. The estimates remain close to the deterministic IMSRG(2) results, while their statistical uncertainty decreases as more independent loops are included. This illustrates the reduction of sampling uncertainty through independent-trajectory averaging. However, a small systematic bias remains at this low walker population, which is analyzed and corrected below.

\revise{To summarize, the above $N_\mathrm{w}$- and $N_\mathrm{loop}$-convergence tests show the statistical convergence of SRGQMC with increasing walker population and number of independent loops. We also noticed small finite-population biases in SRGQMC calculations. To resolve these effects, we perform dedicated high-statistics tests and introduce the replica method to resolve this finite-population bias as below.}

\subsection{Finite-population bias and replica method}
\label{sec:replica_correction}

\revise{With a finite walker population in SRGQMC calculations, the sampled Hamiltonian fluctuates around its ensemble mean. Since the generator also depends on this same stochastic Hamiltonian, the correlation between the generator and the Hamiltonian will induce a bias that cannot be eliminated by averaging over many loops. To make this finite-population bias more clear, we write the IMSRG flow equation as
\begin{equation}
    F(H)=\mathcal{P}_r[\mathcal{L}(H),H],\qquad
    \frac{dH}{ds}=F(H),
    \label{eq:replica_flow_map}
\end{equation}
where $\eta(H)=\mathcal{L}(H)$ and $\mathcal{P}_r$ retains operators up to the normal-ordered $r$-body level. We note that $F$ is not linear in $H$, which means
\begin{equation}
    \mathbb{E}[F(H)]\neq F(\mathbb{E}[H])
\end{equation}
in general, even when each stochastic update samples the instantaneous flow without bias. Increasing the number of loops reduces statistical uncertainty but does not eliminate this systematic difference.

To correct this effect, we use two independent replicas to estimate the fluctuation-induced contribution and an auxiliary operator to track its accumulation. Specifically, we evolve two standard SRGQMC trajectories, $H^{(1)}$ and $H^{(2)}$, with separate random-number streams, and form their mean and half-difference,
\begin{equation}
    \mathcal{M}=\frac{H^{(1)}+H^{(2)}}{2},\qquad
    \mathcal{D}=\frac{H^{(1)}-H^{(2)}}{2}.
    \label{eq:replica_mean_difference}
\end{equation}
The common mean of the Hamiltonian cancels in $\mathcal{D}$, isolating only the difference between the replicas. As derived below, we can estimate the finite-population bias by propagating an auxiliary operator $B$ alongside the replicas according to
\begin{equation}
    \frac{dB}{ds}
    =\mathcal{P}_r\left([\mathcal{L}(\mathcal{M}),B]
    +[\mathcal{L}(B),\mathcal{M}]\right)+2F(\mathcal{D}),\qquad B(0)=0,
    \label{eq:replica_bias_field}
\end{equation}
and the bias-corrected Hamiltonian is
\begin{equation}
    H_{\mathrm{replica}}=\mathcal{M}-B.
    \label{eq:replica_corrected_estimator}
\end{equation}
In this way, the finite-population bias is greatly suppressed for the Hamiltonian. We first demonstrate the effectiveness of this replica method numerically.

\begin{figure*}[t]
    \centering
    \includegraphics[width=0.85\textwidth]{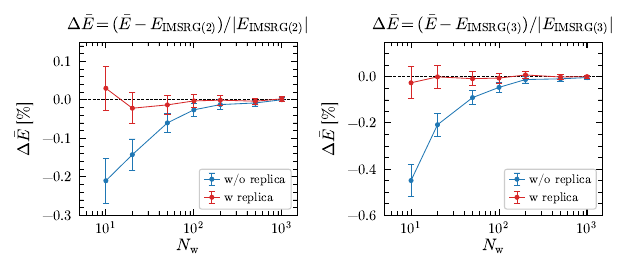}
    \caption{Finite-population bias and replica correction for the Richardson pairing model with $A=4$, $M=8$, $\delta=1\,\mathrm{MeV}$, and $g=0.5\,\mathrm{MeV}$. $M$ is the number of single-particle states. Left and right panels show SRGQMC(2) and SRGQMC(3), respectively. The ordinate is the relative deviation $\Delta\bar{E}$ defined in Eq.~\eqref{eq:replica_relative_bias}, normalized to the magnitude of the deterministic correlation energy. Each point uses $N_\mathrm{loop}=10^4$, with $N_\mathrm{w}$ the total number of walkers of each replica. Error bars denote 95\% confidence intervals of the ensemble mean.}
    \label{fig:bias_replica}
\end{figure*}

Figure~\ref{fig:bias_replica} compares standard and replica-corrected SRGQMC(2) and SRGQMC(3) energies for the pairing model. We use $N_\mathrm{loop}=10^4$ for each walker population. We plot the relative deviation of loop-averaged SRGQMC results with respect to the corresponding deterministic IMSRG results,
\begin{equation}
    \Delta \bar{E}=
    \frac{\overline E_{\mathrm{SRGQMC}}-E_{\mathrm{IMSRG}}}
    {|E_{\mathrm{IMSRG}}|}\times 100\%,
    \label{eq:replica_relative_bias}
\end{equation}
expressed in percent, where all energies are correlation energies. Error bars of $\Delta \bar{E}$ are 95\% confidence intervals.
We can see that the standard SRGQMC results indeed show a systematic deviation at small $N_\mathrm{w}$, and the replica-corrected SRGQMC results show no statistically resolved finite-population bias in these tests, even at the smallest walker number. At $N_\mathrm{w}=10$, the correction reduces the SRGQMC(2) deviation from $-0.210\%$ to $+0.030\%$, and the SRGQMC(3) deviation from $-0.449\%$ to $-0.026\%$, without increasing the statistical uncertainty. All corrected points are consistent with their deterministic references within the displayed confidence intervals. The replica method therefore suppresses the observed finite-population bias at both truncation levels to below the statistical resolution of these tests.

We now explain the origin of the auxiliary equation. Write $H=\mu+\xi$, with $\mu=\mathbb{E}[H]$ and $\mathbb{E}[\xi]=0$. The quadratic structure of $F$ gives the mean-flow equation
\begin{equation}
    \frac{d\mu}{ds}=F(\mu)+S,\qquad S=\mathbb{E}[F(\xi)].
    \label{eq:replica_mean_flow}
\end{equation}
Here $S$ is the additional drift generated by stochastic fluctuations. Let $H_\star$ denote the deterministic solution and $b=\mu-H_\star$ its ensemble bias. Subtracting $dH_\star/ds=F(H_\star)$ and expanding $F(\mu-b)$ yields
\begin{equation}
    \frac{db}{ds}=\mathcal{P}_r\left([\mathcal{L}(\mu),b]
    +[\mathcal{L}(b),\mu]\right)+S-F(b).
    \label{eq:replica_exact_bias}
\end{equation}
This relation identifies both the source of the bias and its subsequent propagation.

For independent, identically distributed replicas, $\mathcal{D}=(\xi^{(1)}-\xi^{(2)})/2$, and we have the identity
\begin{equation}
    2\mathbb{E}[F(\mathcal{D})]
    =\frac{\mathbb{E}[F(\xi^{(1)})]+\mathbb{E}[F(\xi^{(2)})]}{2}
    =S.
    \label{eq:replica_bias_source}
\end{equation}
To obtain the auxiliary equation, we first neglect the term $F(b)$, which is quadratic in the bias and thus a next-order effect. Then we replace the unknown ensemble mean $\mu$ by $\mathcal{M}$, and use $2F(\mathcal{D})$ to sample $S$. Finally we recover the auxiliary equation in Eq.~\eqref{eq:replica_bias_field}. The resulting $B$ estimates the accumulated bias through its linear propagation. Although the identity of Eq.~\eqref{eq:replica_exact_bias} is exact, Eq.~\eqref{eq:replica_bias_field} is an approximation: the omitted quadratic term and correlations between $\mathcal{M}$ and $B$ can still leave a residual bias. However, the good agreement in Fig.~\ref{fig:bias_replica} shows the effectiveness of the correction at the tested statistical precision.

Finally, in the replica method, both parallel trajectories contribute to the energy estimate. At a fixed number of loops, their independent averaging in principle reduces the standard error of the uncorrected mean by $1/\sqrt{2}$ relative to one trajectory per loop. The real additional computational cost is the calculation of the correction term $B$. However we should note that we do not need to use deterministic contractions for this additional calculation, because Eq.~\eqref{eq:replica_bias_field} has a structure similar to that of the standard SRG flow equation and has a linear dependence on $B$. As a result, we can use a similar stochastic algorithm to solve Eq.~\eqref{eq:replica_bias_field} with well-controlled computational cost. This replica method thus provides an efficient way to deal with the finite-population bias in SRGQMC calculations, which should be useful in future applications.}

\section{Scaling Analysis}
\revise{In this section we perform a scaling analysis for SRGQMC, taking SRGQMC(3) as an example. We use the pairing model with $A=4$, $\delta=1\,\mathrm{MeV}$, and $g=0.5\,\mathrm{MeV}$, varying the single-particle basis size over $M=8,\,10,\,12,\,14,$ and 16. We estimate the computational cost required to reach a target accuracy of \(2\%\) relative to the deterministic IMSRG(3) correlation energy, by varying $N_\mathrm{w}$ and $N_\mathrm{loop}$. The replica method is used to correct the finite-population bias in all cases.

Figure~\ref{fig:bias_scaling_srgqmc3} shows the estimated CPU time required to reach the target $2\%$ accuracy. The core CPU time includes generator construction, spawning, annihilation, normalization, and deterministic contractions. The most expensive contractions in a generic dense IMSRG(3) implementation scale as $\mathcal{O}(M^9)$~\cite{PhysRevC.103.044318,PhysRevC.110.044316}. In our standard SRGQMC(3) calculations, fitting $t=aM^n$ gives $n=3.4(2)$, where the parentheses denote one standard error from the fit and $t$ is the CPU time. This much milder observed scaling illustrates the practical benefit of stochastic sampling at fixed accuracy. The fitted exponent shows the effective scaling for the pairing model considered here.

\begin{figure*}[t]
    \centering
    \includegraphics[width=0.45\textwidth]{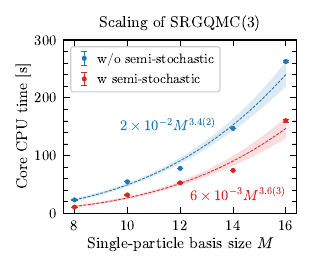}
    \caption{Estimated core computation time at a fixed $2\%$ correlation-energy target accuracy for SRGQMC(3), with and without semi-stochastic treatment, in the pairing model with $A=4$, $\delta=1\,\mathrm{MeV}$, and $g=0.5\,\mathrm{MeV}$. Both variants use the replica method to correct the bias, and the additional replica-correction cost is excluded. Dashed curves are power-law fitting results, and shaded regions indicate pointwise one-standard-error bands of the fitted curves.}
    \label{fig:bias_scaling_srgqmc3}
\end{figure*}

We further test the semi-stochastic variant in which the relatively inexpensive commutator contributions with ranks no higher than two are evaluated deterministically. Stochastic sampling is retained for the remaining higher-order contributions, including the feedback of three-body operators into the lower-rank flow equations. The fitted exponent is $n=3.6(3)$, consistent within fit uncertainties with that of the original scheme. The fitted coefficients are approximately $a=2.1\times10^{-2}\,\mathrm{s}$ and $6.0\times10^{-3}\,\mathrm{s}$ without and with the semi-stochastic treatment, respectively. The estimated semi-stochastic costs are approximately $46\%$--$68\%$ of the original costs over the cases considered, corresponding to a speedup of about $1.5$--$2.2$.

These results support the promise of hybrid deterministic--stochastic evolution for higher-order IMSRG applications. Evaluating the inexpensive low-order contributions exactly reduces sampling noise and lowers the cost of reaching a given accuracy, while retaining stochastic access to the more demanding higher-order contributions. Reduced fluctuations may also mitigate noise-induced finite-population effects, complementing the replica correction.}

\bibliographystyle{apsrev4-2}
\bibliography{reference}